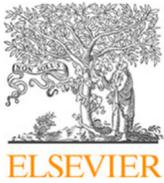

Contents lists available at ScienceDirect

# Applied Ocean Research

journal homepage: www.elsevier.com/locate/apor

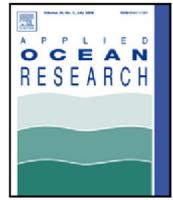

Research paper

# Fully differentiable boundary element solver for hydrodynamic sensitivity analysis of wave-structure interactions

Kapil Khanal [a],*, Carlos A. Michelén Ströfer [b], Matthieu Ancellin [c], Maha N. Haji [d]

[a] *Cornell University, Ithaca, 14850, NY, USA*
[b] *Sandia National Laboratories, Albuquerque, 87123, NM, USA*
[c] *Mews Labs, Paris, France*
[d] *University of Michigan, Ann Arbor, MI, USA*

## ARTICLE INFO

## ABSTRACT

Accurately predicting wave-structure interactions is critical for the effective design and analysis of marine structures. This is typically achieved using solvers that employ the boundary element method (BEM), which relies on linear potential flow theory. Precise estimation of the sensitivity of these interactions is equally important for system-level applications such as design optimization. Current BEM solvers are unable to provide these sensitivities as they do not support automatic differentiation (AD). To address these challenges, we have developed a fully differentiable BEM solver, MarineHydro.jl, for marine hydrodynamics, capable of calculating diffraction and radiation coefficients, and their derivatives with high accuracy. MarineHydro.jl implements both direct and indirect BEM formulations and incorporates two Green's function expressions, offering a trade-off between accuracy and computational speed. Gradients are computed using reverse-mode AD within the Julia programming language. As a first case study, we analyze two identical floating spheres, evaluating gradients with respect to physical dimensions, inter-sphere distance, and wave frequency. Verification studies demonstrate excellent agreement between AD-computed gradients and finite-difference results. In a second case study, we leverage AD-computed gradients to optimize the mechanical power production of a pair of wave energy converters (WECs). This represents the first application of exact gradients obtained from BEM solver in WEC power optimization. Both studies offer valuable insights into hydrodynamic interactions and advance the understanding of layout optimization. Beyond power optimization, the differentiable BEM solver highlights the potential of AD for offshore design studies. It paves the way for broader applications in machine learning integration, optimal control, and uncertainty quantification of hydrodynamic coefficients, offering new directions for advancing wave-structure interaction analysis and system-level optimization.

## 1. Introduction

The determination of the wave-induced response of an offshore structure depends on its hydrodynamic coefficients. These coefficients are determined by considering both the wave diffraction problem (how waves are scattered by a stationary structure) and the wave radiation problem (how waves are generated by the motion of the structure itself). Several methods exist to solve both the wave diffraction and radiation problems, including analytical approaches (Hulme, 1982), and numerical solvers (Ancellin and Dias, 2019; Liu, 2019; Babarit and Delhommeau, 2015). Surrogate models are also created from the numerical and experimental data for cheaper evaluation of the coefficients (Zhang et al., 2020). Analytical and surrogate models, however, are often limited in their applicability due to geometrical assumptions (e.g., axisymmetry of a structure), modeling methods (e.g., a solution being comprised of certain basis functions) or rely on pre-trained data (e.g., in the case of surrogate model). Consequently, numerical solvers remain the preferred choice for general hydrodynamics problems due to their flexibility and accuracy across diverse geometries.

Numerical hydrodynamic solvers generally utilize the boundary element method (BEM) which focuses on the boundaries of the domain rather than the entire volume, thereby significantly reducing computational effort while maintaining the accuracy. This approach transforms the governing partial differential equations (PDEs) into boundary integral equations (BIEs), making it particularly effective for problems with infinite or semi-infinite domains as is the case for analyzing the hydrodynamic forces on offshore bodies. Mature BEM solvers (Ancellin and Dias, 2019; Babarit and Delhommeau, 2015; Lee and Newman, 2003; Liu, 2019) are currently used for the design and analysis of large

---

offshore structures such as, offshore wind turbines (Ashuri et al., 2014), and wave energy converters (WECs) (Teixeira-Duarte et al., 2022).

Recent advancements in computational methods highlight the importance of sensitivity analysis in engineering design (Martins and Kennedy, 2021). Sensitivity analysis elucidates how variations in input parameters (such as ocean wave frequency) influence model outputs (such as electrical power output of a WEC), enabling optimization, uncertainty quantification, and robust design. Automatic differentiation (AD), also known as algorithmic differentiation, is a powerful technique for exact gradient calculation. Exact gradients refer to those computed directly and accurately for the discretized numerical problem, in contrast to approximate gradients obtained through methods such as finite differences or other numerical approximations. AD works by systematically applying the chain rule of differentiation to computer programs, enabling precise and efficient derivative computations. It does so by augmenting the original program with additional code to compute derivatives alongside the original computations (Griewank, 2003; Bartholomew-Biggs et al., 2000). Unlike finite difference methods, AD computes derivatives analytically, ensuring high accuracy and computational efficiency. This technique has been successfully applied in fields such as design optimization (Martins and Ning, 2022), machine learning (Rumelhart et al., 1986), optimal control (Grund, 1985), inverse problems (Tortorelli and Michaleris, 1994), and uncertainty quantification using adjoint-based formulations (Bigoni, 2015). In computational fluid dynamics (CFD), for example, AD has enabled the development of gradient-based optimization algorithms that significantly improve the design of aerodynamic systems (Kenway et al., 2019; Towara and Naumann, 2013). In optimal control, AD has facilitated the precise calculation of control sensitivities, enhancing the performance of control systems in applications ranging from aerospace to robotics (Grund, 1985).

However, these AD-enabled advancements are yet to be widely adopted in marine hydrodynamics. Unlike domains such as acoustics (Takahashi et al., 2022; Silva et al., 2023), structures (Ho Choi and Man Kwak, 1990), and electromagnetics (Koh et al., 1992), where BEM solvers integrate gradient computations, the hydrodynamic kernel function called the *free-surface Green's function* poses numerous numerical difficulties such as singularity for its evaluation and differentiation (Newman, 1985; Liang et al., 2021). While existing marine hydrodynamic BEM solvers are powerful, they are not inherently differentiable, as they lack support for AD, a critical limitation for optimization and sensitivity analysis. Consequently, optimization workflows using traditional BEM solvers have relied either on gradient-free methods, such as statistical (Sobol', 2001) or heuristic (Teixeira-Duarte et al., 2022) algorithms, or on gradient-based techniques that approximate sensitivities using finite difference (Gomes et al., 2012). However, finite difference methods are both computationally expensive and prone to inaccuracies. These limitations make such approaches impractical for large-scale problems involving many interacting bodies, where the cost of repeated gradient evaluations becomes prohibitive. Gradients can be accurately computed in the case of analytical and surrogate model due to availability of the closed form expression. But, analytical methods are constrained by simplifying assumptions, while surrogate models, despite balancing efficiency and accuracy, suffer from limited applicability outside their training domains (Forrester and Keane, 2009).

The lack of accurate and efficient gradients for hydrodynamic coefficients restricts system-level applications, where scalable methods are essential for analyzing subsystems in one coupled model. Multidisciplinary Design Optimization (MDO), a framework that integrates and optimizes across multiple interacting disciplines or subsystems simultaneously, has shown promise in addressing this challenge. Recent studies underscore the advantages of incorporating gradients into MDO frameworks for offshore floating structures (Patryniak et al., 2022). For example, analytical gradient derivations for substructures of offshore floating wind turbines have been integrated with broader analysis, advancing practical engineering capabilities (Dou et al., 2020). Gradients also offer new opportunities for the development of marine renewable energy systems, as emphasized in the U.S. DOE "Next Generation Marine Energy Software Needs Assessment" report (Ruehl et al., 2023), and hold potential for accelerating machine learning-based marine hydrodynamic innovations (Zheng et al., 2024).

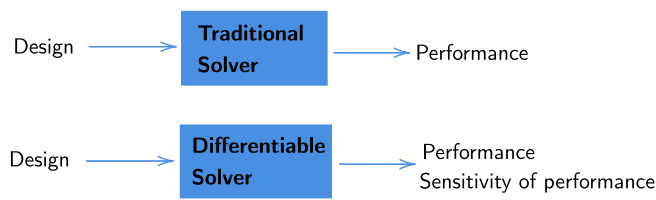

**Fig. 1.** A differentiable solver provides both the performance of the design as well as the sensitivity of the performance to the design parameters.

In contrast to traditional BEM solvers, this paper introduces a new BEM solver–referred to as *MarineHydro.jl* for the remainder of this paper specifically developed to address the limitations outlined above. MarineHydro.jl is a differentiable hydrodynamic solver implemented in Julia (Bezanson et al., 2017), capable of computing not only accurate hydrodynamic coefficients but also their exact gradients with respect to input parameters such as wave frequency, geometry, and degrees of freedom. This enables precise sensitivity analyses and gradient-based optimization, which are otherwise hindered in existing BEM solvers due to the absence of AD. MarineHydro.jl directly propagates derivatives throughout its computational pipeline using AD. This ensures that the gradients reflect the true behavior of the discretized numerical system, without the numerical noise or computational overhead associated with finite differences.

The gradients as computed by MarineHydro.jl are exact, meaning that no approximation was done specifically for computing the hydrodynamic gradients. This is slightly different from saying that we compute the exact analytical gradient of the true Green's function (approximation free). While one of the Green's function used, Wu et al. (2017), is itself an approximation of the exact free-surface Green's function, the gradients computed are exact with respect to this approximation. For the Delhommeau's exact Green's function (Delhommeau, 1987), this solver does calculate the exact Green's function derivative using AD. Both of these Green's functions are included in the solver.

By supporting both direct and indirect boundary integral formulations (Ancellin, 2024) and multiple Green's function models, MarineHydro.jl offers flexibility in balancing computational cost and accuracy across different stages of the design process. Its differentiable architecture makes it well-suited for integration into modern design frameworks such as Multidisciplinary Design Optimization (MDO) and opens new possibilities for advancing the design of offshore systems including wave energy converters and floating wind turbines where gradient-based methods can significantly improve performance, robustness, and development speed. A comparison between differentiable and traditional solvers is illustrated in Fig. 1.

The remainder of this paper is organized as follows. In Section 2, we discuss the formulation of MarineHydro.jl and the underlying Green's function approximation. Following in Section 3, we discuss MarineHydro.jl's adjoint-based differentiation approach as well as detail its implementation in Julia. Section 4 demonstrates MarineHydro.jl's capabilities using two case studies: the first to demonstrate the applicability of computing accurate gradients for two identical heaving floating spheres; and the second to demonstrate the ability of MarineHydro.jl to be used in gradient-based optimization of mechanical power for a pair of identical WECs. We then conclude and outline avenues for future work in Section 5. The code for this paper is available open source at https://github.com/symbiotic-engineering/MarineHydro.jl.

## 2. Formulation and hydrodynamic verification

MarineHydro.jl uses an integral equation formulation based on the free-surface Green's function (Falnes, 2002; Newman, 1985). We refer to the standard textbook Refs. (Falnes, 2002) for detailed description on the wave structure interaction problem, linear water wave theory assumptions, and associated boundary conditions (for infinite depth case). In Section 2.1, we discuss the Green's functions used in MarineHydro.jl.

### 2.1. Evaluation of the Green's function

Numerous approaches have been explored in the literature to formulate and compute the Green's function for the linear potential flow problem (Xie et al., 2018). Developing computationally efficient approximations for the Green's function and its gradients remains an active area and evolving field of research, as highlighted in studies such as Xie et al. (2018), Mackay (2019), John (1950). Newman (1985) outlines efficient numerical algorithms for evaluating the free-surface Green's function and its derivatives for linearized three-dimensional wave motions, considering both infinite and constant finite fluid depths (Newman, 1985, 1986). By leveraging series expansions and multi-dimensional polynomial approximations, Newman's methods significantly enhance the computational efficiency of Green's function calculations compared to traditional numerical integration techniques. These polynomial approximations replace the exact computations, which are often difficult to evaluate and differentiate with the required level of accuracy (Liang et al., 2021). Other approximations of the Green's function utilize computational domain decomposition, such as the six-domain approach in Newman's method (Lee and Newman, 2003), and apply specialized techniques like Legendre or double Chebyshev polynomial approximations for different ranges (Xie et al., 2018).

MarineHydro.jl, developed in this paper, allows users to choose between either an exact expression for the Green's function derived by Delhommeau (1987) or a recently derived global approximation developed by Wu et al. (2017) for the case of infinite depth. The exact expression implementation is significantly slower but is highly accurate and is included for verification. Though we only discuss one approximation for the Green's function in the implementation presented here, the software architecture of MarineHydro.jl enables seamless integration of any Green's function formulation, allowing users to compute hydrodynamic coefficients and their gradients efficiently. This flexibility supports design optimization workflows, permitting the use of faster Green's function approximations during early-stage design studies and transitioning to more precise but computationally intensive options for late-stage analyses, all within the same software framework. This architecture also supports future integration of finite-depth Green's function approximations, particularly those that build upon or reuse infinite-depth solutions.

Fig. 2 summarizes the architecture of MarineHydro.jl and its various operations. MarineHydro.jl calculates the influence matrices (operators) $S$ and $D$ or $K$ (see Appendix C for further details) for a floating body discretized into flat panels. The choice of direct and indirect BIE formulation (see Appendix C for further details) and the choice of the Green's function affects the accuracy and efficiency of hydrodynamic forces calculation. The wave environment parameters are wave frequency ($\omega$) and wave direction ($\beta$).

The remainder of this section will focus on the implementation of the approximation of the Green's function derived by Wu et al. (2017). This Green's function approximation is accurate enough for the determination of accurate hydrodynamic coefficients, and hence sufficient for design optimization studies (Liang et al., 2018).

In this formulation, the Green's function, $G$, is written as

$$4\pi G = -\frac{1}{R} + L + W \tag{1}$$

where $R$ represents the Euclidean distance between the source point $P(x, y, z \le 0)$ and field point $Q(\bar{x}, \bar{y}, \bar{z} \le 0)$, and $L$ and $W$ account for non-oscillatory local flow and pulsating surface waves on the free surface components, respectively. The first term in Eq. (1), $-\frac{1}{R}$, is often referred to as the "Rankine" term. All coordinates and the Green's function are expressed in non-dimensional form, normalized by the wavenumber ($k$).

The wave component $W$ can be expressed as:

$$W(h, v) = 2\pi \left[ H_0(h) + iJ_0(h) \right] e^v \tag{2}$$

where $H_0(h)$ and $J_0(h)$ are the zeroth-order Struve and Bessel functions, respectively, $h = \sqrt{(x-\bar{x})^2 + (y-\bar{y})^2}$ is the horizontal distance and $v = (z + \bar{z}) \le 0$ is the vertical distance (from the free surface) between the panels containing points $P$ and $Q$. The approximations for $H_0(h)$ and $Y_0(h)$ are detailed in Newman (1984).

The local flow component $L$ is approximated as:

$$L \approx \frac{-1}{d} + \frac{2P}{1+d^3} + L' \tag{3}$$

where

$$P \equiv e^v \left( \log \frac{d-v}{2} + \gamma - 2d^2 \right) + d^2 - v \tag{4}$$

and

$$L' \approx (\rho(1-\rho)^3(1-\beta))A(\rho) - \beta B(\rho) - \frac{\alpha C(\rho)}{1 + 6\alpha\rho(1-\rho)} + \beta(1-\beta)D(\rho) \tag{5}$$

where $d = \sqrt{h^2 + v^2}$, $\rho = \frac{d}{1+d}$, $\beta = \frac{h}{d}$, $\gamma$ is Euler's constant, and $A(\rho)$, $B(\rho)$, $C(\rho)$, and $D(\rho)$ are 9th-order polynomials in $\rho$ with coefficients specified in Wu et al. (2017). The first term in (3), $\frac{1}{d}$, is often referred to as the mirror flow or sometime "reflected Rankine" term. Wu et al. (2017) similarly derive an approximation for local flow component ($L^*$) and the wave component ($W_h$) for $\nabla G = [G_x, G_y, G_z]$.

This approximation eliminates the need for complex integral evaluations of $L$ by expressing it as a single polynomial approximation valid across the entire domain, enabling faster computation and simpler implementation through basic polynomial evaluations. The parameters $R, h, v$, and $d$ define the computational domain and characterize the interactions between the source point $P$ and the field point $Q$. In BEM codes, the computational flow domain is primarily determined by $h$ and $v$, representing the horizontal and vertical distances between the points $P$ and $Q$ relative to the free surface.

Several methods outlined in Xie et al. (2018) partition the flow domain into multiple subregions and rely on extensive tabulation and interpolation. In contrast, Wu et al. (2017) formulate the solution using elementary functions that are valid and consistent across the full domain. This not only avoids the additional computational overhead but also ensures that the solution and its derivatives are easily accessible. From a numerical perspective, this method is well-suited for AD due to its inherent simplicity and efficiency. Both forward and backward passes can be parallelized, making this approach computationally efficient and straightforward to implement.

### 2.2. Verification of hydrodynamic coefficients and excitation forces

In this subsection, we verify the accuracy of MarineHydro.jl by comparing its computed hydrodynamic coefficients against analytical solutions for a surface-piercing hemisphere. Specifically, we assess the solver's ability to compute added mass and damping coefficients by referencing the analytical results provided in Hulme (1982). The comparison, detailed in Figs. 3 and 4 for the heave and surge coefficients, respectively, shows good agreement between MarineHydro.jl and the analytical solutions. Any discrepancy in the coefficients compared to analytical results is likely due to the use of a simpler integration method (one point approximation) over the panels. Future work will address this by incorporating more accurate and differentiable integration or quadrature methods.

Similarly, we compute the wave excitation forces for comparison purposes. Fig. 5 presents the comparison of the Froude–Krylov and diffraction forces for a surface piercing unit hemisphere obtained from MarineHydro.jl with those obtained using the exact expression for the

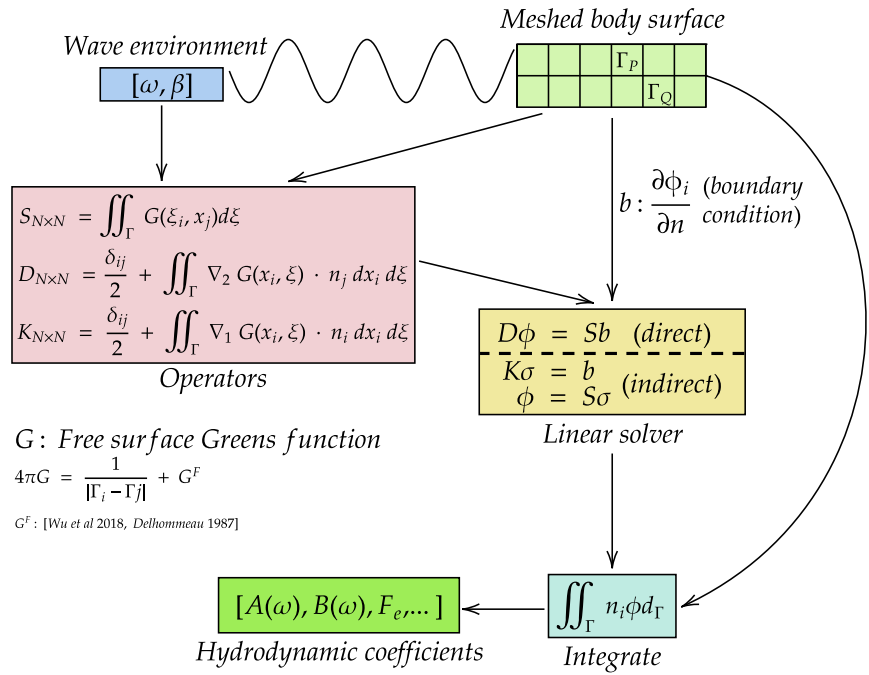

**Fig. 2.** Julia implementation of BEM software. $\omega$ and $\beta$ are the wave frequency and direction, respectively, $D$, $S$, an $K$ are the BEM matrices obtained by evaluating and integrating Green's, $G$, function, and $b$ is the boundary condition. $\phi$ is the hydrodynamic potential and $A$ and $B$ are the hydrodynamic coefficients after integrating $\phi$ over the immersed body surface.

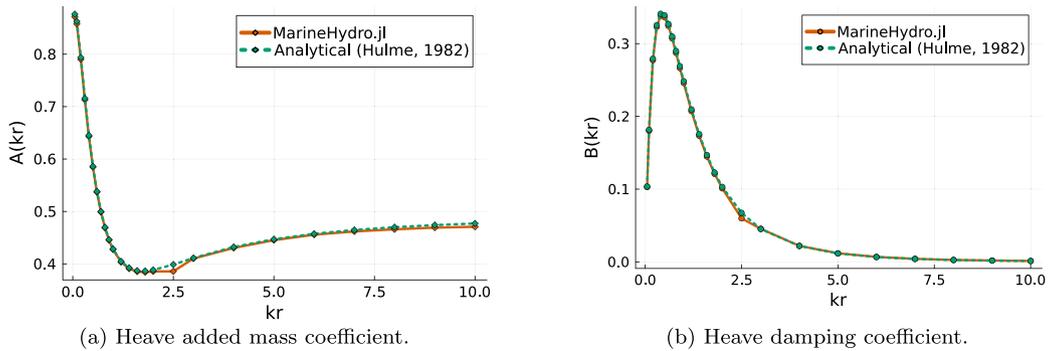

(a) Heave added mass coefficient.

(b) Heave damping coefficient.

**Fig. 3.** Comparison of the non-dimensional heave hydrodynamic coefficients for (a) added mass and (b) damping, as computed by the MarineHydro.jl, with analytical (Hulme, 1982) results for a surface piercing unit hemisphere. Added mass is non-dimensionalized as $\frac{A}{\frac{2}{3}\rho\pi r^3}$ and damping as $\frac{B}{\frac{2}{3}\rho\omega\pi r^3}$.

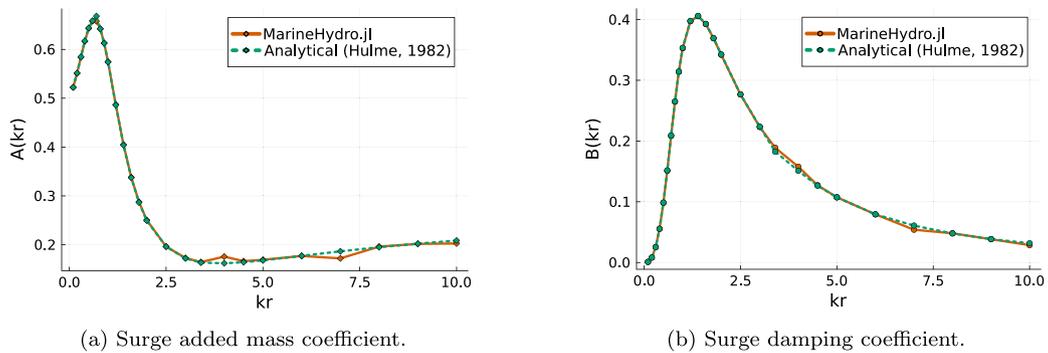

(a) Surge added mass coefficient.

(b) Surge damping coefficient.

**Fig. 4.** Comparison of non-dimensional surge hydrodyamic coefficients for (a) added mass and (b) damping, as computed by the MarineHydro.jl, with and analytical results for a surface piercing unit hemisphere from Hulme (1982). Added mass is non-dimensionalized as $\frac{A}{\frac{2}{3}\rho\pi r^3}$ and damping as $\frac{B}{\frac{2}{3}\rho\omega\pi r^3}$.

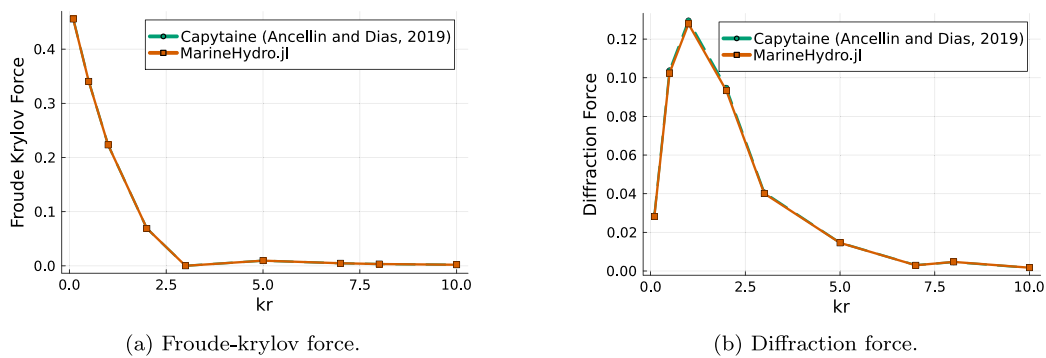

(a) Froude-krylov force.

(b) Diffraction force.

**Fig. 5.** Comparison of the non-dimensional heave excitation force (a) Froude–Krylov component and (b) diffraction component, as computed by the MarineHydro.jl, with BEM solver Capytaine (Ancellin and Dias, 2019) for a surface piercing unit hemisphere. Froude–Krylov and Diffraction forces are normalized by $\pi \rho g H r^2$, where H is wave height and r is the radius.

**Table 1**
Comparison of open-source solvers based on their capabilities.

| Solver | Green's function method | | Formulation | Irreg. Freq. removal | Parallelization | Differentiable |
|---|---|---|---|---|---|---|
| | Wu et al. (2017) (Wu et al., 2017) | Delhommeau (1987) (Delhommeau, 1987) | | | | |
| MarineHydro.jl v0.1.0 | Yes | Yes | Direct & Indirect | No | No | Yes |
| Capytaine v2.2.1 (Ancellin and Dias, 2019) | No | Yes | Direct & Indirect | Yes | Yes | No |
| HAMS (Liu, 2019) | Yes | No | Direct only | Yes | Yes | No |

Green's function derived by Delhommeau (1987) as implemented in the BEM software Capytaine (Ancellin and Dias, 2019). This comparison highlights the solver's accuracy in calculating wave excitation forces, a critical component for assessing wave-structure interactions.

MarineHydro.jl is sufficiently accurate for practical use in both diffraction and radiation problems, particularly in early-stage design and sensitivity analyses. The deviation observed in these comparisons at some frequencies may be attributed to the influence of irregular frequencies, a known challenge in such simulations, and will be investigated further in future work. Irregular frequencies arise when the numerical solution of the BIEs becomes non-unique at certain discrete frequencies. These correspond to non-trivial solutions of the homogeneous equations (Fredholm of the second kind) used in Green's function formulations for surface-piercing bodies, leading to either non-unique or undefined solutions (Lee and Sclavounos, 1989). Although they do not reflect any physical phenomenon, irregular frequencies can significantly distort the computed hydrodynamic coefficients and their gradients. Therefore, any method used to suppress or eliminate these frequencies (Liu and Falzarano, 2017; Burton and Miller, 1971; Lee and Sclavounos, 1989) must also preserve the differentiability of the solver to ensure gradient propagation remains accurate.

### 2.3. Solver capability comparison

MarineHydro.jl integrates multiple capabilities for hydrodynamic coefficient calculation including both direct and indirect BIE formulations (see Appendix C for further details of these formulations). Table 1 compares the current capabilities of MarineHydro.jl against other open-source solvers.

In the next section, we discuss the differentiation of MarineHydro.jl and review concepts relevant to developing the differentiable code.

## 3. Differentiability of the BEM solver

A naive way to obtain the gradients of the computed hydrodynamic coefficients and excitation forces is to numerically approximate them using finite differences. However, this method involves solving the hydrodynamic problem multiple times — once for each design variable — resulting in a computational cost of $O(m \times n^3)$ for the case of the direct

BIE solver, where $m$ is the number of design variables and $n$ represents the panels in the floating body. This makes the approach computationally expensive for high-dimensional problems. Alternatively, AD can be used to compute the gradient of complex numerical code, especially in cases where manual derivation, symbolic differentiation, or finite difference approximation are challenging or impractical.

The required Jacobian ($J_{M \times N}$) of the $M$ hydrodynamic coefficients with respect to $N$ input parameters (e.g., wave frequency, body geometry variables, etc.) can be constructed either column-by-column or row-by-row. These two approaches correspond to two AD modes: Forward mode (also called pushforward) and reverse mode (also called pullback, backpropagation, or adjoint mode). The forward mode operation computes the Jacobian-vector product (JVP), which propagates an input perturbation ($\partial x$) through the function to compute the resulting perturbation in the output ($\partial y$):

$$\partial y = J \cdot \partial x. \tag{6}$$

This approach is useful for building the Jacobian column-by-column, as it calculates the effect of a small change in each input on the outputs.

The backward mode computes the vector-Jacobian product (VJP) by introducing a perturbation in the output space ($\partial y$) and propagating it back through the function to determine the impact on the input space ($\partial x$):

$$\partial x = J^\top \cdot \partial y. \tag{7}$$

This method is ideal for building the Jacobian row-by-row as it computes how small changes in each output affect all inputs simultaneously. It is particularly efficient when dealing with optimization objectives or constraints where only a few outputs matter (Organization, 2024; Griewank and Walther, 2008).

Although the solver can perform both forward- and reverse-mode AD, we adopt the reverse-mode formulation (adjoint mode) throughout this work because most design-optimization problems involve many design variables but only one or a few objective functions. Following Jameson's pioneering application to transonic-aircraft design in 1988 (Jameson, 1988)—itself an outgrowth of Pironneau's optimal-control ideas (Grund, 1985), discrete adjoint have become a mainstay of high-dimensional optimization (Giles and Pierce, 2000). Contemporary surveys detail the adjoint-based method and how it is applied to decouple the cost of sensitivity computation from the number of design variables across different applications (Allaire, 2015; Lettermann et al., 2024; Mader et al., 2008).

Traditional BEM solvers have not prioritized gradients, resulting in separate performant solvers and user-facing post-processing tools, which complicates integration with modern differentiation-based methods. While tools like Tapenade (Hascoet and Pascual, 2013) and other algorithmic differentiation frameworks (Elliott, 2018) facilitate differentiability through techniques such as operator overloading or program transformations, these methods are often ill-suited for BEM implementations that comprise legacy codes and a mix of programming languages. Transforming BEM code (using algorithmic tools like Tapenade) as it is currently written can lead to inefficiencies; the forward solver's design, including aspects like parallelization strategies and memory management, plays a critical role in determining the efficiency and accuracy of the differentiability (Towara and Naumann, 2013). Moreover, these methods often require significant modifications or entirely new solvers, as demonstrated by Rohrer and Bachynski-Polić (2024), whose workaround using adjoint solver in OpenMDAO (Gray et al., 2019) still faced memory limitations and inefficiencies for large-scale design tasks. These approaches often fix the mesh resolution and skip differentiating the influence matrix assembly process, and instead perform finite-difference approximations of the matrix entries, limiting their accuracy and ability to adapt to design changes. This fixed-matrix approach is not ideal for optimization tasks, as highlighted by Rohrer and Bachynski-Polić (2024), who emphasized the need for solvers inherently designed for differentiability. Accuracy and efficiency challenges arise in practical applications, such as WEC layout optimization, where multiple linear solves may be required for analyses involving multiple interacting bodies and an additional adjoint solve is required for each linear solve.

MarineHydro.jl described in this paper makes differentiability a core feature. MarineHydro.jl implements AD through the Green's function evaluation and its integrals that assemble the influence matrix, enabling exact gradient computation under any mesh refinement — this is unlike conventional approaches that numerically finite-difference gradients using precomputed entries of an influence matrix. This provides flexibility for large-scale design optimization studies where an optimizer may make significant design adjustments and mesh refinement may be necessary. The novel contribution of MarineHydro.jl is its ability to combine adjoint formulations with AD to compute exact gradients, eliminating the need for numerical approximations.

### 3.1. Deriving the adjoint

Two primary approaches exist for formulating adjoints in the context of flow control and optimization (PDE-constrained optimization) problems: optimize-then-discretize (OtD) (Gunzburger, 1987) and discretize-then-optimize (DtO). The OtD approach formulates the optimization problem (using the continuous equations) and derives its discrete form afterward, while the DtO approach first discretizes the governing equations and then derives the optimization problem, often resulting in different numerical behavior. Nadarajah and Jameson (2012) (Nadarajah and Jameson, 2000) review these approaches in the context of aerodynamic design problems and Bradley (2010) also discusses them in detail for general setting (Bradley, 2010). While both methods theoretically yield equivalent results, practical considerations favor one over the other depending on the specific problem (Bradley, 2010).

The OtD approach involves deriving the unconstrained optimization problem in its continuous form by formulating the Lagrangian:

$$\mathcal{L}(\theta, u, \lambda) = \mathcal{J}(\theta, u) + \lambda_i^T \mathcal{F}(\theta, u), \tag{8}$$

where $\theta$ are the design variables, $u$ are the state variables, $\mathcal{J}(\theta, u)$ is the objective function, $\mathcal{F}(\theta, u)$ represents the constraints (e.g., boundary conditions or flow field equations), and $\lambda_i$ are the Lagrange multipliers used to enforce each of these constraints. The stationarity condition for optimality, $\nabla_u \mathcal{L} = 0$, provides the sensitivity of the boundary conditions, $F(\theta, u)$, with respect to the state variables $u$. Solving the resulting Karush-Kuhn–Tucker (KKT) conditions, discretized as a linear system, yields the solution to the adjoint boundary value problem (Ragab, 2004). Ragab (2004) applied the OtD approach to free-surface flows, implementing a time-domain panel code for hydrodynamic ship design under forward speed in waves using translating and pulsating free-surface Green's function. The work formulated a wave resistance functional and a target pressure distribution for ship surfaces. They modified the panel code for the adjoint free-surface condition. However, the gradients derived in this approach were restricted to the specific objective function and may deviate from numerical results after discretization, as observed by Ragab (2004) who found up to 6% error in the gradients when compared with finite differences (Ragab, 2004). This could be due to discretization inconsistencies between the two solvers.

In contrast, MarineHydro.jl described in this paper adopts the DtO approach, also known as the discrete adjoint method. This method is preferred for working with numerical solvers as it directly uses the discretized equations ensuring that the computed gradients are consistent with the numerical scheme and approximations used in forward solution. This consistency between the computed function value and its gradients is crucial for reliable optimization, particularly when the objective function depends on these values. It is worth noting that these gradients differ from analytical gradients derived from the continuous adjoint formulation. This distinction arises from the use of the discrete adjoint method, which enables the solver's differentiability for the numerical problem being solved. This is a deliberate feature of our methodology and not a limitation of the solver itself.

Consider minimizing $\mathcal{J}(\phi(\theta), \theta)$ with respect to $\theta$, where $\phi(\theta)$ is defined implicitly by the equation

$$D(\theta)\phi(\theta) - S(\theta)b(\theta) = 0. \tag{9}$$

where $D$ and $S$ are $n \times n$ asymmetric complex-valued dense square matrices where $n$ is the number of panels. $\mathcal{J}$ is the objective function dependent on the potentials on each panel, $\phi$, and the design variables, $\theta$.

The total derivative of $\mathcal{J}$ with respect to $\theta$ is expressed as:

$$\frac{d\mathcal{J}}{d\theta} = \frac{\partial \mathcal{J}}{\partial \theta} + \left(\frac{\partial \mathcal{J}}{\partial \phi}\right)\frac{\partial \phi}{\partial \theta}. \tag{10}$$

where implicit differentiation gives

$$\frac{\partial D}{\partial \theta}\phi + D\frac{\partial \phi}{\partial \theta} - \frac{\partial S}{\partial \theta}b - S\frac{\partial b}{\partial \theta} = 0, \tag{11}$$

and therefore

$$\frac{\partial \phi}{\partial \theta} = D^{-1}\left(S\frac{\partial b}{\partial \theta} + \frac{\partial S}{\partial \theta}b - \frac{\partial D}{\partial \theta}\phi\right). \tag{12}$$

An additional linear system can be formed to solve for the adjoint variable ($\lambda$) by letting:

$$\lambda^T D = \frac{\partial \mathcal{J}}{\partial \phi}. \tag{13}$$

The gradient of $\mathcal{J}$ with respect to $\theta$ in Eq. (10) can then be expressed as:

$$\frac{d\mathcal{J}}{d\theta} = \frac{\partial \mathcal{J}}{\partial \theta} + \lambda^T\left(S\frac{\partial b}{\partial \theta} + \frac{\partial S}{\partial \theta}b - \frac{\partial D}{\partial \theta}\phi\right), \tag{14}$$

where the expensive partial derivatives $\frac{\partial \phi}{\partial \theta}$ is skipped. The total derivative $\frac{d\mathcal{J}}{d\theta}$ efficiently combines contributions from the boundary condition $b$ and its perturbation $\frac{\partial b}{\partial \theta}$, the influence matrix $S$, matrix perturbations $\frac{\partial S}{\partial \theta}$ and $\frac{\partial D}{\partial \theta}$, and the solution vector $\phi$. Similar derivation through the method of Lagrange multipliers can be done and is included in the Appendix F for interested readers.

In Eq. (10), $\frac{\partial \mathcal{J}}{\partial \theta} = 0$ in most applications, though for applications where there may be direct dependence of objective function to inputs, it can be calculated using AD. A similar derivation from the indirect BIE could also be performed to obtain the indirect BIE adjoint linear system.

To compute the partial derivatives required in the adjoint equation, Eq. (14), various methods can be utilized, including finite differences, complex-step differentiation, symbolic differentiation, or AD. However, the accuracy and efficiency of gradient computation depend heavily on the selected approach.

Finite differences, for example, can suffer from inefficiencies and truncation errors. This method approximates the derivative as:

$$f'(x) \approx \frac{f(x+h) - f(x)}{h}, \tag{15}$$

where $h$ represents a small perturbation in each input. Each input $x$ perturbation requires two function $f$ evaluations, and this process must be repeated for all inputs thus this method scales with number of inputs. For BEM solvers, this approach becomes computationally expensive due to the high cost of evaluating the Green's function and solving the discretized boundary integral equation.

Complex-step differentiation reduces the truncation error; however it requires the real and imaginary values be separable and does not improve upon the efficiency of finite differences (Martins and Ning, 2022). This requirement may necessitate modifications to the algorithms and Green's function implementation in the computer code to support complex number inputs. Symbolic differentiation is exact but may be cumbersome to derive especially for computer programs. Out of these, AD is exact and scalable and is thus used in this solver. Differentiation is performed with respect to input variables which can be panel properties (e.g., vertices and centers), the dimensions of floating bodies, or the ocean wave frequency.

As illustrated in the derivation above, the adjoint method reduces the number of the linear solves to just two, regardless of the number of the design variables (Allaire, 2015). This is equivalent to solving an additional adjoint boundary value problem for a single scalar function output. The computational cost, therefore, decreases from $m \times O(n^3)$ to $2 \times O(n^3)$, where $m$ is the number of design variables, and $n$ is the size of the dense, asymmetric matrices. Furthermore, for a factorization-based solver such as LU decomposition, the factorization ($A = LU$) can be reused for both forward and adjoint solve and the complexity will be just $O(n^3)$. Additionally, the matrix required for the adjoint solver is the transpose of the matrix from the forward solver. This ensures that the matrix's conditioning remains consistent, eliminating the need for higher precision in the Green's function evaluations in the adjoint solve.

### 3.2. Automating adjoint derivation

Manually deriving and assembling the adjoint equation can be complex and error prone. For example, in WEC design optimization, evaluating the objective function may require repeatedly solving the state equation along with additional linear systems, such as the equations of motion. A solver that abstracts this process can greatly enhance usability and streamline the design workflow. Reverse-mode AD is an effective alternative, assembling the partial derivatives required for the adjoint equation without explicitly solving for adjoint variables. This approach works by propagating output sensitivities backward through the computational chain. Unlike forward-mode AD, which traces input perturbations, reverse-mode AD is better suited for high-dimensional problems involving scalar objectives (Elliott, 2018). Notably, reverse-mode AD and the adjoint method are conceptually analogous.

Differentiable BEM solvers have two choices of applying AD *through* a solver algorithm (which is generally not efficient) or using AD *with* a solver algorithm (i.e., doing an explicit solve to apply the implicit function theorem but using AD to get the terms that form the linear system for the solve). We chose the latter implementation. Relying on AD to differentiate through linear solvers (sometime iterative solvers like GMRES), can introduce inefficiencies and errors (Griewank and

Walther, 2008). In implicit differentiation, for a system, $F(x, u) = 0$, where $u$ represents the solution and $x$ is a design variable, gradients are obtained by solving a linear system at optimality, $F(x, u(x)) = 0$. This bypasses the need for AD to "unroll" external iterative solver computations. We utilize libraries like ImplicitAD (Ning and McDonnell, 2023) and Zygote.jl (Innes, 2018) to implicitly differentiate through the linear solver, allowing for efficient and automated adjoint setup within a single differentiable pipeline. This approach is solver-agnostic, meaning users can adopt any state-of-the-art linear solver without modification, ensuring compatibility with a wide range of Julia-based scientific computing tools.

In our solver implementation, in addition to automating the adjoint of the linear solver, we differentiate the Green's function, which presents singularity challenges (Liang et al., 2021). In the case of the Rankine and reflected Rankine terms (mirror effect), which are both of the form $\frac{1}{x}$, $x \to 0$ results in singularities. These are handled using a Rankine integration algorithm described by Newman (1986), with AD applied directly to compute derivatives of the integral efficiently. Similarly, in case of integration for the exact expression by Delhommeau (Delhommeau, 1987), the differentiable HCubature method is adopted available in Julia package Integrals (SciML, 2025). We apply AD to the midpoint quadrature method for integrating the Green's function, noting that the accuracy of the results is influenced by the chosen integration method. This method is employed for constant panels, offering a balance between simplicity and computational efficiency.

The frequency-dependent Green's function depends on the horizontal ($h$) and vertical ($v$) distances between panels relative to the free surface. However, spatial derivatives can become problematic when $d = \sqrt{h^2 + v^2} \to 0$ (Liang et al., 2021). The derivative of $d$ with respect to $h$ or $v$ is undefined at $(0, 0)$, as the limits of the gradient differ: $\lim_{h \to 0} \nabla d(h, 0) \neq \lim_{v \to 0} \nabla d(0, v)$. This issue is mitigated by introducing a small regularization constant ($d + \epsilon$), ensuring gradient consistency with finite difference approximations. Regularization, relaxation, and reparameterization are common techniques in differentiable codes to stabilize gradients and ensure differentiability almost everywhere, making programs robust and fully differentiable (Blondel and Roulet, 2024). We can employ either an approximate or exact Green's function in the forward solve; thus its gradients should align consistently with the discretized forward solution to ensure the solver's reliability and accuracy in design optimization studies. Liang et al. (2021) highlighted discrepancies between direct differentiation of approximated Green's functions and analytically derived results. To assess the accuracy of our approach, we compare the gradients of hydrodynamic coefficients computed via AD with those obtained through finite difference approximations. The results show excellent agreement with numerical approximation via finite difference. This serves as extension study of the validation by Wu et al. (2018) (Liang et al., 2018), who demonstrated that the approximated Green's function (Wu et al., 2017) is sufficiently accurate for design applications. Similarly, these gradients are sufficient for practical design optimization studies where sensitivities of coefficients are of importance.

### 3.3. Verification of the Green's function gradients and radiation force derivatives

To verify the performance of MarineHydro.jl, we evaluate its ability to compute the gradients of the Green's function as well as the gradients of the radiation forces on a surface piercing unit hemisphere using the adjoint formulation with AD as compared to those obtained through the finite-difference method. There exists a difference in accuracy of the two components of the Green's function: wave and Rankine. The wave component of the Green's function (the second term in Eq. (1) which is frequency dependent) is an approximation that results in reduced accuracy compared to exact analytical values. The wave component can be viewed as effect from the free surface as discussed in Section 2.1. Consequently, its gradients may not perfectly align with analytical results. In contrast, the gradients of the Rankine component (the first term in Eq. (1) which is frequency independent) can be evaluated exactly. Thus, we can directly compare against the analytical added mass of a submerged sphere resulting from just the Rankine component of Green's function. This is equivalent to added mass of a submerged sphere in still water which does not include a free surface term. Specifically, the gradient for the added mass of a sphere in surge in still water (and no free surface), given by

$$\frac{1}{A}\frac{\partial A}{\partial r} = \frac{2\pi\rho r^2}{\frac{2}{3}\pi r^3\rho},\tag{16}$$

is compared with the result obtained using AD, as illustrated in Fig. 6. As we can see from the figure, the two methods align well, indicating the accuracy of the AD implementation in MarineHydro.jl. The accuracy is expected to improve and converge toward analytical results with an increase in the number of panels.

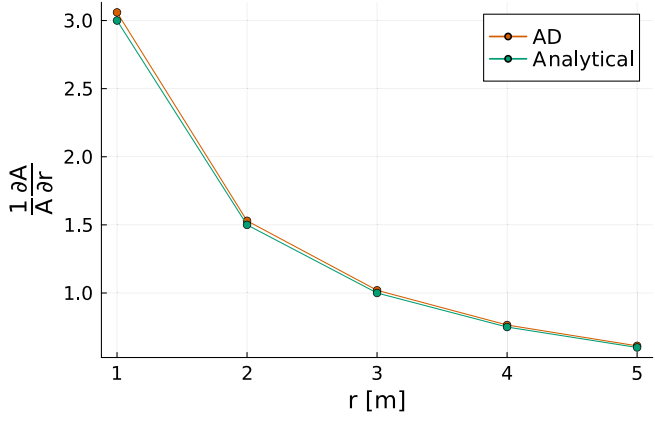

**Fig. 6.** Comparison of the analytical and AD-computed normalized gradients $\left(\frac{1}{A}\frac{\partial A}{\partial r}\right)$ of added mass ($A$) for a surging hemisphere sphere in still water shows that the AD computed sensitivities match the analytical results.

From the velocity potential over each body's immersed surface (at each panel), the first order radiation force is computed for body$_k$ and body$_l$ due to their motion is computed based on (Kashiwagi et al., 2005):

$$\mathbf{F}_{ij}^{(k)(l)} = \omega^2 \mathbf{A}_{ij}^{(k)(l)} - i\omega\mathbf{B}_{ij}^{(k)(l)} = -\rho \iint_{S_k} \mathbf{\Phi}_j^l \, \mathbf{n}_i^k \, dS \tag{17}$$

where $\mathbf{A}_{ij}^{(k)(l)}$ is the matrix of added mass coefficients, where each element represents the added mass for the $i$th degree of freedom of the $k$th body due to the $j$th mode of motion of the $l$th body and similarly $\mathbf{B}_{ij}^{(k)(l)}$ is the matrix of radiation damping coefficients. $\mathbf{\Phi}_j^l$ is the velocity potential field for the $j$th mode of the $l$th body. $\mathbf{n}_i^k$ is the generalized normal vector for body$_k$. Note that, this only provides the contribution to this generalized force component related to each mode of the bodies.

For the verification test case, we only consider one heaving body; thus, the coefficients $A$ and $B$ are scalar values and obtained using MarineHydro.jl. In this context, let $f : \mathbb{R} \to \mathbb{C}$ map the dimension of the floating device ($\theta$) to the radiation force. Then, $f$ can be differentiated by treating it as $f : \mathbb{R} \to \mathbb{R}^2$, where the real and imaginary parts are differentiated independently. To compute the derivative of the radiation force $F_{R_i}$ in the $i$th degree of freedom with respect to the design variable $\theta$, the real and imaginary components of the complex valued $F_{R_i}$ are differentiated separately given in Eq. (17), as follows:

$$\frac{\partial F_{R_i}}{\partial \theta} = \frac{\partial \Re F_{R_i}}{\partial \theta} + i\frac{\partial \Im F_{R_i}}{\partial \theta} \tag{18}$$

where $\Re F_{R_i}$ and $\Im F_{R_i}$ are the real and imaginary parts of the radiation force, respectively. By Eq. (17), this effectively amounts to differentiating each of the hydrodynamic coefficients ($A_{ij}^{(k)(l)}$ and $B_{ij}^{(k)(l)}$) separately for interacting bodies.

Figs. 7 and 8 demonstrate that these real and imaginary gradients as computed from MarineHydro.jl are exact when compared to those computed with finite differences, confirming their accuracy. In Figs. 7(a) and 7(b), we take the gradient with respect to the radius of the sphere, $r$, for a fixed incoming wave frequency $\omega = 1.03$ rad/s. We see that for both added mass and damping in heave, the gradient computed using AD with respect to radius for a range of hemisphere radii (1 m to 5 m) is nearly the same as that computed using finite difference, with an absolute error of $\leq 10^{-7}$ kg/m in added mass and $\leq 10^{-7}$ Ns/m in damping.

Figs. 8(a) and 8(b) show the gradient with respect to the incident wave frequency, $\omega$, for a fixed sphere radius, $r = 1$ m. We see that for both added mass and damping in heave, the gradient computed using AD for a range of incident wave frequencies (0 rad/s to 3.6 rad/s) is nearly the same as that computed using finite difference, with an absolute error of $\leq 10^{-7}$ kg/m in added mass and $\leq 10^{-7}$ Ns/m in damping.

Similarly, MarineHydro.jl can calculate the sensitivity in the case of either the direct and indirect BIE formulations, with a comparison for added mass and damping shown in Fig. C.18 in Appendix C. Note that these sensitivities are as accurate as they can get given the imprecise nature of the finite difference method. Thus, the AD computed sensitivities are accurate for the practical purposes.

In the next subsection, we delve into the implementation details of the solver.

### 3.4. Implementation for differentiability

MarineHydro.jl is implemented in the Julia programming language and makes differentiability a core feature. This approach allows AD tools like Zygote to "natively" differentiate the source code via source-to-source AD. The implementation primarily rewrites the existing Fortran-based Green's function (Liu, 2019; Wu et al., 2017; Ancellin and Dias, 2019) and integral algorithms, adopting functional programming paradigms where appropriate. Zygote.jl supports reverse-mode differentiation, which is crucial for design optimization, the primary motivation for this solver. Forward-mode differentiation, enabled by Julia's dual number support, is also available and can be extended by implementing forward chain rules (as outlined in the following Section 3.5), though reverse-mode remains the focus for optimization tasks.

When a user needs to calculate the gradient of an output such as power, which depends on the BEM solution, this solver automatically records each step of the solution process. It decomposes the process into elementary operations, computes the analytical derivatives for these operations, and reassembles them to obtain the full gradient. For the linear solve, the solver sets up an adjoint equation using the computed gradients and solves it for gradient evaluation with respect to the input variables. This is unlike the existing approach by Rohrer and Bachynski-Polić (2024) where Jacobian of the influence matrices are estimated using finite differences in the adjoint equation. Since the Green's function itself is approximated and has some errors when directly differentiating (Liang et al., 2021), the finite difference will compound such errors. Additionally, their implementation approach is memory bound due to the need to estimate and store large Jacobian matrices. The approach and architecture of MarineHydro.jl is scalable and computes exact analytical derivatives. Efficient Green's function approximation and its derivatives can be included and differentiated with ease because of the differentiation rules (chain rules) that can be provided to the solver.

The adjoint solver in MarineHydro.jl, constructed using an AD engine like Zygote.jl, is automatically generated when a user requests gradients for specific inputs. During the backward solve (reverse-mode or adjoint-mode), the solver propagates sensitivities back through the computational graph from the output to the inputs. This approach

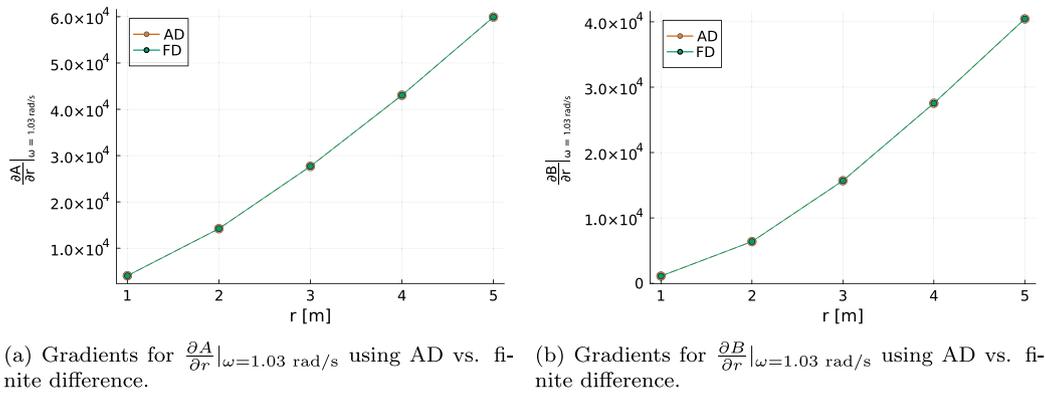

(a) Gradients for $\frac{\partial A}{\partial r}|_{\omega=1.03 \text{ rad/s}}$ using AD vs. finite difference.

(b) Gradients for $\frac{\partial B}{\partial r}|_{\omega=1.03 \text{ rad/s}}$ using AD vs. finite difference.

**Fig. 7.** Comparison of gradients computed by AD to those computed by finite difference (FD) for (a) heave added mass $\frac{\partial A}{\partial r}|_{\omega=1.03 \text{ rad/s}}$ (b) and heave damping $\frac{\partial B}{\partial r}|_{\omega=1.03 \text{ rad/s}}$. Note that added mass is non-dimensionalized as $\frac{A}{2/3 \rho \pi r^3}$ and damping as $\frac{B}{2/3 \rho \omega \pi r^3}$.

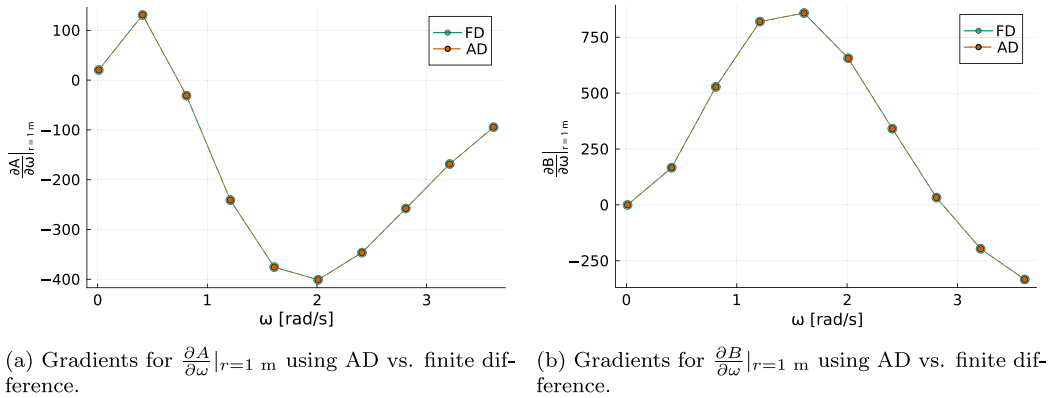

(a) Gradients for $\frac{\partial A}{\partial \omega}|_{r=1 \text{ m}}$ using AD vs. finite difference.

(b) Gradients for $\frac{\partial B}{\partial \omega}|_{r=1 \text{ m}}$ using AD vs. finite difference.

**Fig. 8.** Comparison of gradients computed by AD to those computed by finite difference (FD) for (a) heave added mass $\frac{\partial A}{\partial \omega}|_{r=1 \text{ m}}$ (b) and heave damping $\frac{\partial B}{\partial \omega}|_{r=1 \text{ m}}$. Note that added mass is non-dimensionalized as $\frac{A}{2/3 \rho \pi r^3}$ and damping as $\frac{B}{2/3 \rho \omega \pi r^3}$.

scales with the number of outputs, not inputs, making it highly efficient for problems with many input variables.

Notably, while MarineHydro.jl's interface aligns with familiar solvers (Ancellin and Dias, 2019), it uniquely automates the backward solve process as shown in Fig. 9. Unlike many state-of-the-art solvers that rely on users to implement computationally intensive numerical methods, such as finite differences, MarineHydro.jl simplifies gradient evaluation, requiring minimal effort from the user. By employing the adjoint state method and AD, it seamlessly integrates the adjoint solver alongside the forward solver.

Although the current implementation is not yet fully optimized for memory efficiency, it is designed to overcome the constraints of earlier methods and provide high accuracy for realistic, complex design tasks.

In the following subsection, we outline the methodology for enabling and propagating gradients throughout the BEM source code.

### 3.5. Custom chain rules for automatic differentiation

For use cases where MarineHydro.jl may need to be used with other external codes, such as a tool that generates a mesh from geometric parameters, custom differentiation rules can be augmented within MarineHydro.jl. For code that is not written in Julia or contains unsupported constructs, however, we implement custom gradient propagation rules using the ChainRulesCore.jl package (White et al., 2024). For example, for the case studies in Sections 4.1 and 4.4, we leverage the external libraries Capytaine (Ancellin and Dias, 2019) for meshing and hydrostatics, which are not differentiable. We therefore use ChainRulesCore.jl to specify custom chain rules to propagate gradients through the AD computational graph and calculate the gradients with respect to the dimensions. Future implementations of MarineHydro.jl will aim to replace external meshing and hydrostatics routines with fully differentiable mesh pre-processing code such that studies involving hydrodynamics are automatically differentiable from geometry creation to coefficient calculations.

Mesh pre-processing involves generating the mesh vertices and panel properties for the BEM solver. For example, a meshing function $f_{\text{mesh}}(r) \rightarrow$ vertices calculates the mesh vertices for a sphere given the radius $r$. The Jacobian of this function is then computed using finite differences and augmented with a custom reverse-mode rule for AD in Zygote.jl, enabling gradient propagation. The reverse-mode differentiation rule for Zygote.jl computes the vector-Jacobian product, $\frac{\partial f_{\text{mesh}}}{\partial r}^\top \cdot \delta y$, where $\delta y$ is the gradient propagated back from the output of BEM solver to the meshing function.

This approach can be extended to other mesh panel properties required for BEM coefficient calculations. Although finite differences are used for mesh pre-processing in MarineHydro.jl, this remains computationally acceptable since these operations are far less computationally intensive than the hydrodynamic analysis. By defining custom AD rules, MarineHydro.jl ensures compatibility with differentiable pipelines while maintaining accuracy in numerical gradients. Algorithm 1 in Appendix D illustrates the implementation process for the custom reverse-mode differentiation rule for mesh computations used in MarineHydro.jl. The custom AD rules developed here are not limited to mesh pre-processing but can be extended to any external CAD geometry tools, enabling a fully differentiable pipeline for calculating hydrodynamics coefficients. This approach can also integrate post-processing operations and non-computationally intensive tasks in hydrodynamic analysis. Non-computationally intensive tasks refer to those that do not scale with the number of input variables, as opposed to operations that scale polynomial like linear solvers, which are more resource intensive.

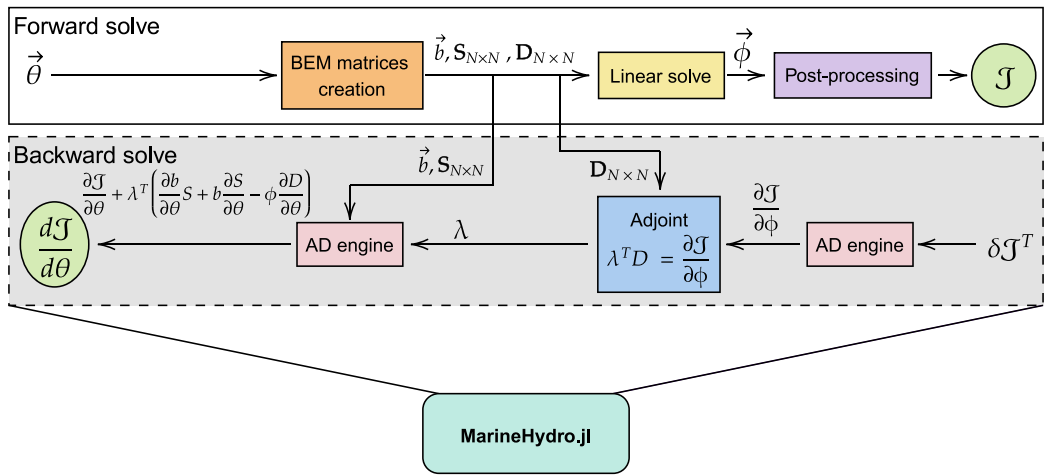

**Fig. 9.** Forward and backward solve workflow of a differentiable BEM solver (MarineHydro.jl): Forward solve computes the hydrodynamic coefficients, while backward solve employs adjoint and automatic differentiation (AD) for sensitivities. Backward solve is also referred as adjoint solve or reverse-mode automatic differentiation.

**Table 2**
Computational cost of Green's function integrals and influence matrix sensitivities with respect to the panel $z$-coordinate using reverse-mode AD engine Zygote.jl (Innes, 2018). Exact calculations are performed between panels for comparison (although approximation can be done for panels that are far away from each other). These measurements were obtained using Julia (version 1.10.5) running on a Linux platform with a single CPU thread and 256 GB memory.

| Green function | Integral type | Coefficient (time) | Gradient (time) |
|---|---|---|---|
| Rankine | S | 1747.45 ns | 768.737 μs |
| Rankine | D | 3558.83 ns | 6.533 ms |
| Wu (Wu et al., 2017) | S | 74.9424 ns | 37.200 μs |
| Wu (Wu et al., 2017) | D | 238.953 ns | 523.291 μs |
| Exact Delhommeau (Delhommeau, 1987) | S | 138472 ns | 1.889 ms |
| Exact Delhommeau (Delhommeau, 1987) | D | 283603 ns | 11.210 ms |

Reverse-mode AD, while powerful, requires more memory than the forward mode because it caches all intermediates value during the forward solve for use in backpropagation. Future iterations of MarineHydro.jl will explore performance optimizations such as checkpointing and the incorporation of analytical derivatives to reduce memory usage and enhance efficiency (Griewank and Walther, 2008). Additionally, other AD engines available in Julia will be evaluated to identify opportunities for improved performance and scalability.

### 3.6. Solver computational speed

The per-panel performance of MarineHydro.jl, for both matrix assembly and gradient computation using each Green's function, is summarized in Table 2.

Note that the current implementation, while functional, is not yet optimized for either coefficient or gradient calculations. Performance optimization is beyond the scope of this paper, but future work will focus on improving the efficiency of both the forward and gradient computations.

In the following sections, we conduct two case studies to demonstrate some of the capabilities of MarineHydro.jl.

## 4. Application of AD computed sensitivities

In this section, we employ the differentiable capability of Marine-Hydro.jl to compute the sensitivities of hydrodynamic coefficients with respect to their dimension and wave parameters. MarineHydro.jl makes it straightforward to compute these sensitivities without the need for finite-difference approximation.

### 4.1. Case study I: Sensitivity of hydrodynamic coefficients for two identical floating spheres

Exploring the hydrodynamic interactions between closely spaced floating bodies can provide valuable insights for designing systems that leverage constructive wave interaction effects for power generation (Konispoliatis and Mavrakos, 2016; Singh and Babarit, 2014). Analyzing these interactions can be used to identify a separation distance at which simpler models, such as the plane wave approximation (PWA) (Singh and Babarit, 2013), can be applied to significantly reduce the computational costs associated with computing the hydrodynamic response of floating bodies in a given wave environment (Zhang et al., 2022). Currently, this distance is considered to be approximately five times the characteristics dimension of the floating body, beyond which simpler hydrodynamic coefficients approximations can be used with minimal accuracy loss (Singh and Babarit, 2013).

For WECs, understanding how variables such as body dimensions, inter-body spacing, and wave climate influence energy absorption is essential for designing robust and cost-effective array layouts. These factors play a critical role in optimizing power production and ensuring the effectiveness of energy farms (Borgarino et al., 2012).

In this case study, we examine two identical point-absorbers shown in Fig. 10 using MarineHydro.jl described earlier. Each of the point-absorbers are considered as one degree of freedom systems, moving in heave only. Through this case study, MarineHydro.jl is extended to compute two-body hydrodynamic interactions and their sensitivities using adjoint and AD.

For each combination of radius and wavenumber, only one forward solution (BEM solution) and a single gradient calculation is required. We calculate the local sensitivities of each sphere's hydrodynamic added mass and damping coefficients relative to the influence on other sphere. These sensitivities are evaluated with respect to their separation distance $x$ and the frequency of the wave environment $\omega$ for a fixed-size spheres of radius $r$. The dimensionless parameters $kr$ and $\frac{x}{r}$ are varied to calculate the gradients providing insights into hydrodynamic interactions between floating bodies.

The coupled and symmetric added mass matrix (**A**) is given by:

$$\begin{bmatrix} A_{11} & A_{12} \\ A_{21} & A_{22} \end{bmatrix},$$

and the damping matrix (**B**) is:

$$\begin{bmatrix} B_{11} & B_{12} \\ B_{21} & B_{22} \end{bmatrix}.$$

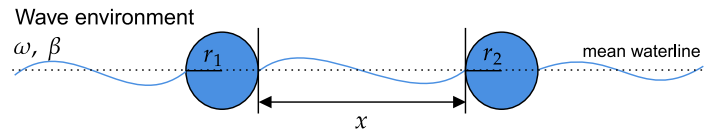

**Fig. 10.** Schematic of a pair of identical floating spheres used for Case Study I.

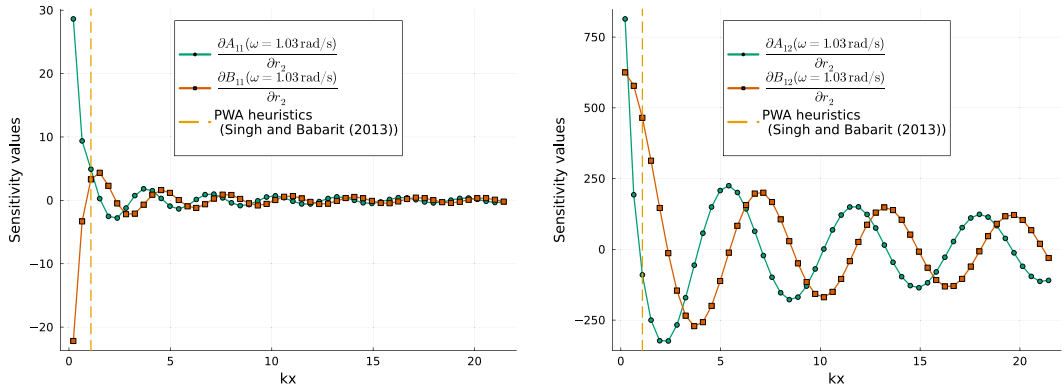

(a) Sensitivity of sphere 1's added mass and damping coefficients with respect to the dimension of sphere 2 as well as the PWA heuristics for simplified interaction modeling (Singh and Babarit, 2013).

(b) Sensitivity of interaction added mass and damping coefficients (off-diagonal) with respect to the dimension of sphere 2 as well as the PWA heuristics for simplified interaction modeling (Singh and Babarit, 2013).

**Fig. 11.** Comparison of sensitivities of added mass and damping coefficients.

are computed for the two heaving spheres. Sensitivities are analyzed to understand interactions, with gradients computed using AD in Marine-Hydro.jl via the exact Green's function expression by Delhommeau (1987).

### 4.2. Sensitivity results and observations

The sensitivity of the added mass $A_{11}$ and damping $B_{11}$ for a unit hemisphere ($r_1 = 1$ m) at $\omega = 1.03$ rad/s, and for a nearby sphere of radius $r_2$, with respect to their dimensions can be used to determine an interaction cutoff range for the PWA (Singh and Babarit, 2013). This helps justify the use of simplified interaction models at sufficiently large separations. Based on numerical studies (Singh and Babarit, 2013), the cutoff distance is typically taken to be five times the characteristic dimension of the bodies involved. For the flow problem considered here, this dimension is taken to be the diameter ($d = 2$ m) of the hemisphere.

Fig. 11(a) illustrates $\frac{\partial A_{11}}{\partial r_2}$ and $\frac{\partial B_{11}}{\partial r_2}$ approaching zero as the separation distance increases, indicating that sphere 1's coefficients become independent of sphere 2's dimension. This result aligns with expectations, as the added mass and damping of a body are influenced by fluid flow around the body, which depends on the geometry of nearby obstacles such as the sea bed or other floating bodies. For the infinite-depth case considered in this solver, the sea bed is not a nearby obstacle. Similar numerical studies can be conducted for various geometries before implementing simplified models at large distances. For instance, Singh and Babarit (2013) used a distance of five times the characteristic dimension as a cutoff to build an approximation model for hydrodynamic interaction. They discuss the application of the PWA in the context of wave interaction within arrays of wave energy converters. Their numerical experiment showed that at such separation distance, the PWA showed reliable results for wave periods ranging from 4–15 s. This is based on the observation that the curvature of outgoing waves diminishes at such distance making them resemble plane waves.

From Fig. 11(a) it seems that this is a good guideline for the diagonal elements of the added mass and damping matrices ($A_{11}$ and

$B_{11}$, respectively), however the same cannot be said for the off-diagonal $A_{12}$ and $B_{12}$ terms. The sensitivity of $A_{11}$ and $B_{11}$, representing self-interaction, tend to stabilize and show minimal sensitivity to variations in separating distance (at large distances). This aligns with the assumptions of the PWA. Fig. 11(b) shows that the sensitivity of the interaction coefficients is oscillatory with the separation distance. This oscillatory sensitivity arises from the constructive and destructive interference patterns in the wave field. Similarly, the sensitivities, as expected, also have a $\frac{\pi}{2}$ phase difference between them. It is also noteworthy that the rates at which the sensitivities of added mass and damping decrease to zero differ, reflecting the distinct hydrodynamic influences on these coefficients.

Therefore, while the PWA serves as a useful tool for simplified hydrodynamic analysis at far distances for self-interaction effects, more analysis is necessary when the precise mutual interaction effects (off-diagonal terms) are critical. These effects are influenced significantly by the relative positioning of the floating bodies in the wave field. In these cases, BEM sensitivity analysis using the AD computed gradients as presented here is recommended.

Figs. 12 and 13 illustrate the sensitivities of the coupled added mass $A_{12}$ and coupled damping $B_{12}$ coefficients, respectively, with respect to the dimension of the pair of identical spheres $r$, swept over the separation distance ratio $\frac{x}{r}$ and non-dimensional radius $kr$. In Figs. 12 and 13, we are calculating the sensitivities of identical spheres, essentially perturbing both radii at the same time and calculating sensitivities across different wave conditions. This study mirrors the layout and design optimization studies where identical WECs are sized and placed in different wave conditions (Teixeira-Duarte et al., 2022). Fig. 12 shows that the sensitivity of $A_{12}$ varies from normalized values 0 to 1 (corresponding to 420.4 kg/m to −750 kg/m dimensionally) showing strong spatial dependency. Negative-to-positive gradient transitions highlight regions where increasing $\omega$ significantly alters the coupled effect, likely due to wave interference.

Similarly, Fig. 13 shows that the damping sensitivities with respect to radius range from normalized values 0 to 1 (corresponding to 2148.89 Ns/m to −1291.43 Ns/m dimensionally). The damping coefficients exhibit regions of increasing or decreasing sensitivity depending on the distance and sphere size. Both added mass and damping sensitivities demonstrate non-uniform variations with separation distance and frequency. Note that, since the irregular frequencies have not yet been removed in MarineHydro.jl yet, the coefficients and their sensitivities may exhibit some accuracy issues at some of those frequencies. However, the variations exist across the frequency range and thus, this study underscores the importance of layout optimization. For instance, in WEC arrays where the dimensions of identical WECs typically change during optimization iterations.

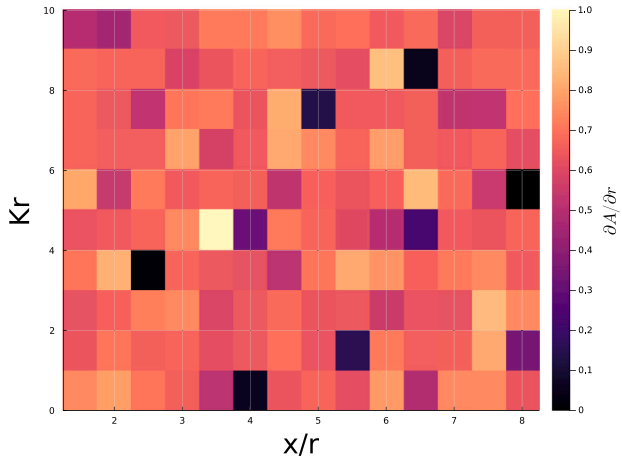

**Fig. 12.** Sensitivity of added mass ($A_{12}$) with respect to radius ($r$) for unit spheres: $\frac{\partial A_{12}}{\partial r}\Big|_{r=1}$ m.

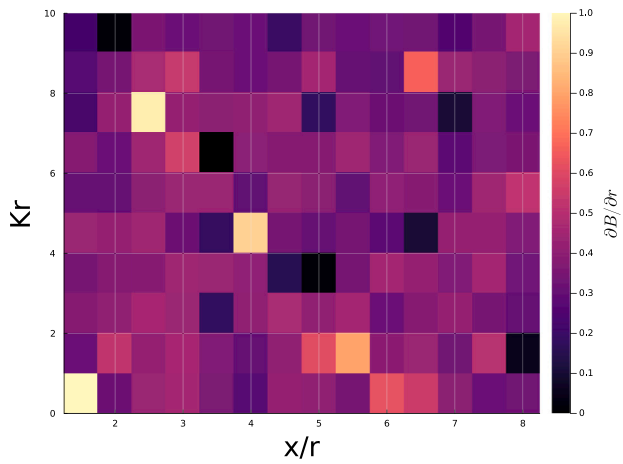

**Fig. 13.** Sensitivity of damping ($B_{12}$) with respect to radius ($r$) for unit spheres: $\frac{\partial B_{12}}{\partial r}\Big|_{r=1}$ m.

Thus, the AD capabilities of MarineHydro.jl can be used for calculating the sensitivities for any kind of geometry, essentially providing a direct and easy approach to the analysis performed by Singh and Babarit (2013) to gain intuition for design studies.

### 4.3. Implications for design and optimization

These results reveal critical "transition regions" in sensitivities, emphasizing the need to account for realistic wave climates (encompassing all relevant $\omega$) in layout optimization studies. For instance, certain changes in sphere dimensions can offset sensitivity changes due to increased separation distances. Ringwood (2025) reviewed a number of design aspects including separation distance and wave scenarios essential for the design optimization of WECs and their controls. With this case study, we confirm that the sensitivities show various non-intuitive interactions. This interplay exists across different frequencies and thus should be exploited in optimization tasks using optimization algorithms to achieve desired performance metrics, such as maximizing energy absorption.

The spatial sensitivities computed using the approximation by Wu et al. (2017) align within 2% of those derived from the exact Delhommeau expressions (Delhommeau, 1987) as shown in Appendix B, Fig. B.17, verifying the solver's accuracy for design optimization tasks when using the faster Green's function approximation. These exact sensitivities are essential for guiding experimental setups and improving the practicality of large-scale numerical design optimization.

### 4.4. Case study II: Optimization of mechanical power from pair of point absorbers

In this section, we demonstrate the capabilities of MarineHydro.jl by optimizing the mechanical power of two identical point absorbers WECs heaving independently for a fixed wave frequency $\omega = 1.03$ rad/s as shown in Fig. 14. The objective is to identify the optimal device dimensions and spacing between the absorbers that maximize mechanical power. While realistic wave spectra, wave roses, and larger WEC arrays are typically considered in practical applications, this example is intentionally simplified to highlight the solver's functionality.

### 4.5. Methodology

To propagate the gradients efficiently, the dynamics, control systems, and power calculations for the WECs were implemented in Julia, with hydrostatics made differentiable using custom chain rules as described in Appendix D. A resistive controller is used, setting the power take-off (PTO) stiffness to zero ($k_i = 0$), and the PTO damping coefficient is set to be equal to the hydrodynamic damping coefficient $\vec{d} = \mathrm{diag}(\mathbf{B})$ for each sphere.

For linear PTO devices, the time-averaged power, P, for each WEC$_i$ is given by:

$$P_i = \frac{1}{2} d_i \left| j\omega \mathbb{E}_i(i\omega) \right|^2, \tag{19}$$

where $\mathbb{E}_i(i\omega)$ is the complex amplitude of heave motion. The coupled dynamics of the system are solved using the frequency-domain linear equation:

$$\vec{\mathbb{E}}(i\omega) = \left[-\omega^2(\mathbf{M} + \mathbf{A}) - i\omega(\mathbf{B} + \mathrm{diag}(\vec{d})) + \mathbf{C} + \mathrm{diag}(\vec{k})\right]^{-1} \vec{\mathbb{F}}(i\omega), \tag{20}$$

where $\mathbf{A}$ is the $2 \times 2$ added mass matrix, $\mathbf{B}$ is the $2 \times 2$ hydrodynamic damping matrix, $\vec{\mathbb{F}}(j\omega)$ is the $1 \times 2$ wave excitation force vector, $\mathbf{C}$ is the $2 \times 2$ hydrostatic stiffness matrix, $\mathbf{M}$ is the $2 \times 2$ diagonal mass matrix, $\mathrm{diag}(\vec{d})$ is a diagonal matrix of the PTO damping coefficients for unit wave amplitude. The hydrodynamic coefficients and wave exciting force ($\mathbf{A}, \mathbf{B}, \vec{\mathbb{F}}(i\omega)$) are computed using MarineHydro.jl, while $\mathbf{C}$ is obtained from Capytaine (Ancellin and Dias, 2019) and integrated into the differentiable pipeline via custom chain rules as described in Section 3.5. Note that, $\mathbf{C}$ is not a function of frequency $\omega$. Although geometry and hydrostatic gradients are estimated using finite differences, this process is still automated by the AD engine (using custom reverse rules), requiring no manual assembly by the user. Custom rules allow the user to provide the derivative rule when AD engines encounter code external to them such as geometry creation. This allows users to use any tool they prefer for geometry and mesh creation. For this study, the focus is on hydrodynamics, as it varies with frequency $\omega$. Hydrodynamic coefficients involve significantly more computationally intensive operations, such as Green's function evaluation, integration, and linear solving, making an efficient method for hydrodynamic sensitivities critical to the design optimization studies.

MarineHydro.jl automates the differentiation of both the BIEs and the equations of motion. This automation eliminates the need for manual adjoint derivations when additional post-processing tasks, such as computing Response Amplitude Operators (RAOs) are part of the sensitivity calculations. For instance, in scenarios where WECs interact across multiple degrees of freedom, the discretized BIE must be solved for each degree of freedom. To demonstrate MarineHydro.jl's scalability, consider a WEC farm with $N$ interacting bodies, each moving in one degree of freedom, and $W$ wave frequencies $\omega$ in a wave spectrum. For this analysis, the computation involves solving $W$ influence matrices, each of size scaling with $N \times N$. These matrices account for interactions between the $N$ bodies and are computed using a fixed number of panels per body. The matrices are to be generated for both radiation and diffraction problems. Similarly, the computational cost for nearby body configuration is higher due to the cost of Green's function evaluation while for bodies far away, simplified approximation is used. Thus, the cost for some configuration will be higher than other for both forward and adjoint solve for same number of panels, interacting bodies and wave frequencies.

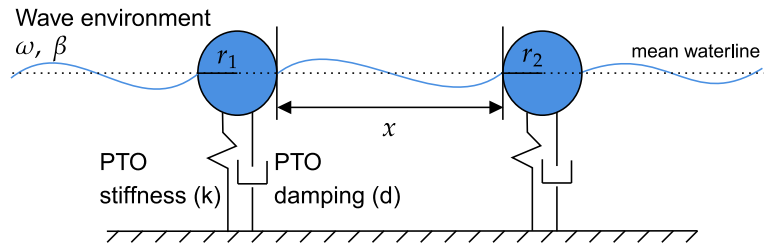

**Fig. 14.** Pair of point absorbers separated by distance $x$.

**Table 3**
Comparison of gradients: Finite difference vs. AD.

| Design variable | Finite difference gradient | AD gradient | Absolute error |
|---|---|---|---|
| radius (1 m) | 170.437647446 | 170.43764737445 | $\leq 10^{-9}$ |
| separation distance (5 m) | $-0.2635335396$ | $-0.2635335398$ | $\leq 10^{-9}$ |

The automated handling provided by MarineHydro.jl makes such complex scenarios more manageable and computationally feasible.

The optimization problem for this case study is formulated as follows:

$$\text{Minimize} \quad \mathcal{J}(\hat{\theta}) = \frac{\sum_i^n P_i}{\frac{4}{3}\pi r^3}, \tag{21}$$

$$\text{by varying} \quad \hat{\theta} = [r, x], \tag{22}$$

$$\text{subject to} \quad x > 0, \tag{23}$$

$$\text{while solving} \quad D(\theta; \omega)\phi - S(\theta; \omega)b = 0, \tag{24}$$

$$-\omega^2\left(\mathbf{M}(\theta) + \mathbf{A}(\phi, \theta)\right) - i\omega\left(\mathbf{B}(\phi) + \text{diag}(\mathbf{B}(\vec{\phi}))\right)$$
$$+ \mathbf{C}(\theta)\vec{\mathbb{E}}(i\omega) - \vec{\mathbb{F}}(\theta)(i\omega) = 0, \tag{25}$$

where $\mathcal{J}$ is the total mechanical power per unit volume, $\hat{\theta}$ is vector of design variables, $r$ is the radius of the point absorber, $x$ is the separation distance, $S$ and $D$ are influence matrices, $b$ is the boundary condition defined in (A.3) and (A.4), and $n = 2$ is the number of interacting WECs. Fig. 15 showcase a typical adjoint based design optimization using the new differentiable solver.

A detailed step-by-step comparison of the computations using the adjoint method and finite difference in this design optimization study is provided in Appendix G.

### 4.6. Results and discussion

The accuracy of the gradients of the objective function with respect to the design variables computed by the MarineHydro.jl was verified against finite difference results. Table 3 shows excellent agreement between the methods.

Table 3 also shows that, at the chosen ocean wave frequency ($\omega = 1.03$ rad/s), the sensitivity of the power calculation for two point

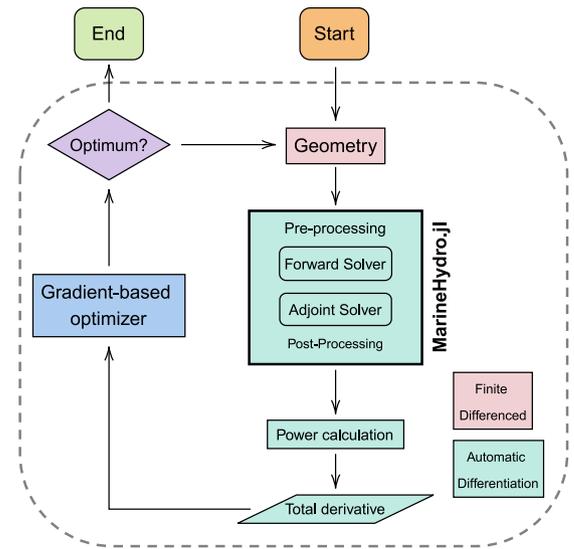

**Fig. 15.** Adjoint based design optimization using the new differentiable solver.

absorbers depends more on their body dimension than their separation distance.

To see how MarineHydro.jl scales with additional design variables, the optimization problem was modified to include some "dummy" design variables which have no effect on the objective function value. Fig. 16 compares the computation times for gradients obtained via numerical finite differences and AD in MarineHydro.jl. It demonstrates that as the number of design variables increases, the time required for a single gradient evaluation using finite difference grows significantly. In contrast, AD maintains a constant computation time regardless of the number of design variables. The current AD implementation exhibits higher computation times, reflected in the elevated $y$-intercept, due to unoptimized reverse-mode differentiation, and this cost increases with mesh resolution. Future versions of MarineHydro.jl aim to significantly reduce these runtimes through targeted performance optimizations.

The optimization was carried out using the Limited-memory Broyden–Fletcher–Goldfarb–Shanno (L-BFGS) algorithm via Optim.jl (Mogensen et al., 2024). The nominal and optimal design variables as well as the optimal objective function value are summarized in Table 4. The optimizer converged in three iterations and 857 s, selecting point absorbers with the smallest allowed radius and maximum separation distance when using AD.

This case study demonstrates the scalability of AD for high-dimensional problems, as the cost of gradient computation remains constant regardless of the number of design variables. Although this case study only considers two design variables, real-world WEC layout optimization typically involves numerous interacting bodies across a range of wave frequencies within a spectrum (Teixeira-Duarte et al., 2022). Importantly, AD's computational efficiency ensures that gradient calculations are equally cost-effective for both small-scale and large-scale optimization tasks, making it well-suited for complex, high-dimensional scenarios.

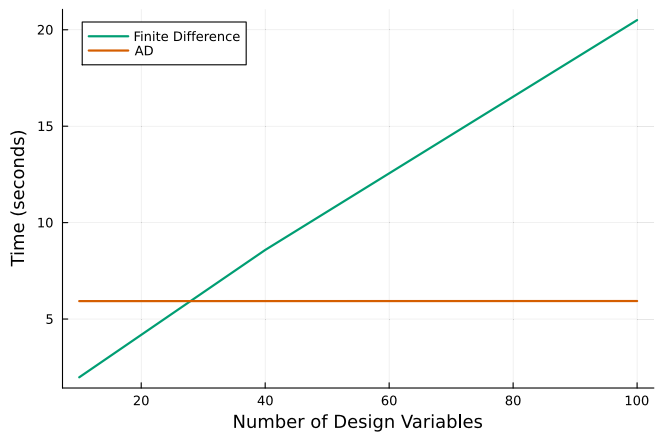

**Fig. 16.** Comparison of computation times: finite difference vs. AD as the number of design variables increase.

**Table 4**
Summary of optimization results.

| Variable | Bounds | Nominal value | Optimal value |
|---|---|---|---|
| Radius ($r$ [m]) | [1.0, 4.0] | 2.0 | 1.0 |
| Separation Distance ($x$ [m]) | [1.0, 4.0] | 2.0 | 3.9 |
| Optimal Objective Function AD | 24.48 W/m³ | | |

The current example, with just two variables, is not ideal for showcasing the full potential of reverse-mode AD but serves as a simplified demonstration. Reverse-mode AD is particularly advantageous for high-dimensional problems, such as the design optimization of multiple interacting bodies in wind-wave layout configurations. For realistic applications, detailed optimization would require greater computational resources, but using this solver with gradient based optimizers is expected to incur significantly lower costs compared to traditional methods like finite differences or heuristic optimization.

## 5. Conclusion and future work

In this work, we developed and implemented a novel fully differentiable BEM solver capable of accurately computing hydrodynamic coefficients and their gradient calculations, known as MarineHydro.jl. By leveraging adjoint and AD, MarineHydro.jl achieves precise and scalable gradient calculations, overcoming limitations in traditional BEM solvers that often requires researchers to rely on finite differences or heuristic methods. The solver also incorporates both exact and approximate Green's functions and supports direct and indirect BIE formulations, thus improving the design workflow by allowing faster Green's function approximations during early-stage design studies and transitioning to more precise but computationally intensive options for late-stage analyses.

Through rigorous numerical experiments, we verified MarineHydro.jl's accuracy against analytical benchmarks and demonstrated its practical utility in two case studies. The first case study analyzed hydrodynamic interactions between two identical floating spheres, revealing critical insights into the sensitivity of coupled hydrodynamic coefficients to design and environmental parameters. The second case study utilized the sensitivity to optimize the mechanical power production of a WEC array, illustrating the potential of MarineHydro.jl for system-level design optimization.

This work highlights the potential of integrating differentiable code into marine hydrodynamics. By eliminating the need for manual adjoint derivation and enabling seamless gradient propagation, the proposed solver simplifies workflows and extends the capabilities of traditional BEM approaches. This advancement opens avenues for integrating sensitivity analysis, machine learning and data-driven approaches in offshore engineering challenges such as WEC farm layout optimization, floating wind turbine design, and other applications. As the offshore industry shifts toward more complex and integrated design frameworks, the adoption of differentiable tools like MarineHydro.jl will be crucial for advancing engineering innovation and operational efficiency.

Future research will aim to enhance MarineHydro.jl's efficiency for both forward computations and backward gradient propagation, optimizing performance on both CPU and GPU architectures. In addition to performance improvements, we aim to extend MarineHydro.jl's application to large-scale systems engineering challenges, such as design optimization and uncertainty quantification in wave-structure interaction analyses. The extension of this solver for finite depth scenario will also be considered in future iterations. Finally, we plan to continue developing the package to support the mesh pre-processing and hydrostatics components in Julia (which currently relies externally on Capytaine), improving the modularity to handle multiple bodies, ensuring a fully integrated and differentiable workflow.

## CRediT authorship contribution statement

**Kapil Khanal:** Writing – review & editing, Writing – original draft, Visualization, Validation, Software, Methodology, Investigation, Formal analysis, Data curation, Conceptualization. **Carlos A. Michelén Ströfer:** Writing – review & editing, Supervision, Methodology, Funding acquisition, Conceptualization. **Matthieu Ancellin:** Writing – review & editing, Software, Methodology. **Maha N. Haji:** Writing – review & editing, Supervision, Resources, Funding acquisition, Conceptualization.

## Funding statement

Sandia National Laboratories is a multi-mission laboratory managed and operated by National Technology & Engineering Solutions of Sandia, LLC (NTESS), a wholly owned subsidiary of Honeywell International Inc., for the U.S. Department of Energy's National Nuclear Security Administration (DOE/NNSA) under contract DE-NA0003525. This written work is authored by an employee of NTESS. The employee, not NTESS, owns the right, title and interest in and to the written work and is responsible for its contents. Any subjective views or opinions that might be expressed in the written work do not necessarily represent the views of the U.S. Government. The publisher acknowledges that the U.S. Government retains a non-exclusive, paid-up, irrevocable, worldwide license to publish or reproduce the published form of this written work or allow others to do so, for U.S. Government purposes. The DOE will provide public access to results of federally sponsored research in accordance with the DOE Public Access Plan.

## Declaration of competing interest

The authors declare that they have no known competing financial interests or personal relationships that could have appeared to influence the work reported in this paper.

## Acknowledgments

We sincerely thank the anonymous reviewers for their valuable comments and constructive suggestions, which have significantly enhanced the clarity of the manuscript. Authors would also like to thank Prof. David Bindel, Dr. John Jasa, Olivia Vitale, Collin Treacy, and Julia discourse members for their valuable feedback and discussion on the simulation, analysis and code derivation in this manuscript. This work was supported in part by the Seedling and Sapling program by the U.S. Department of Energy.

## Appendix A. Direct and indirect boundary integral equations

To solve the Laplace equation efficiently using BEM, the problem is reformulated as a BIE. Following Ancellin (2024), two primary methods for this transformation are discussed and summarized in the following sections: the direct and indirect BIE. Reference textbooks on this topic include Sauter and Schwab (2011) and Gaul et al. (2003).

### A.1. Direct boundary integral equations

The direct BIE, often referred to as the "potential" or "sources-and-dipoles" formulation, is widely used in existing BEM solvers WAMIT (Lee and Newman, 2003) and HAMS (Liu, 2019). Denoting the immersed boundary of the body as $S_B$, the BIE reads

$$\frac{\Phi(x)}{2} + \iint_{S_B} \Phi(\xi)\nabla_2 G(x,\xi)\cdot n(\xi)d\xi = \iint_{S_B} \frac{\partial\Phi}{\partial n}(\xi)G(x,\xi)d\xi \quad \forall x \in S_B \tag{A.1}$$

where $G(x,\xi)$ is the fundamental solution (the Green's function) of the governing Laplace equation, $n$ is the normal vector, and $\nabla_2$ indicates the gradient with respect to the second variable in Green's function, $\xi$ in this case. Using a collocation method and discretizing the boundary into $N$ panels, the equation becomes:

$$D\Phi = S\frac{\partial\Phi}{\partial n} \tag{A.2}$$

where the matrix $S$, sometimes called single-layer operator, is defined as

$$S_{ij} = \iint_{S_{B_j}} G(x_i,\xi)d\xi \tag{A.3}$$

and the $D$ matrix, sometimes referred to as the double-layer operator, is defined as

$$D_{ij} = \frac{\delta_{ij}}{2} + \iint_{S_{B_j}} \nabla_2 G(x_i,\xi)\cdot n_j d\xi \tag{A.4}$$

where $\delta_{ij}$ is the Kronecker delta and $n_j$ denotes the normal vector of panel $S_{B_j}$.

### A.2. Indirect boundary integral equations

One commonly used indirect BIE approach is the "sources" formulation, which introduces a new scalar field, $\sigma$, defined on $S_B$. The equations for $\Phi(x)$ are formulated as:

$$\Phi(x) = \iint_{S_B} \sigma(\xi)G(x,\xi)d\xi \quad \forall x \in S_B \tag{A.5}$$

and

$$\frac{\partial\Phi}{\partial n}(x) = \frac{\sigma}{2} + \iint_{S_B} \sigma(\xi)\nabla_1 G(x,\xi)\cdot n(x)d\xi \quad \forall x \in S_B \tag{A.6}$$

where $\nabla_1$ denotes the gradient taken with respect to the first variable ($x$ in this case).

After discretization as in the direct BIE, these equations are expressed as:

$$\Phi = S\sigma \tag{A.7}$$

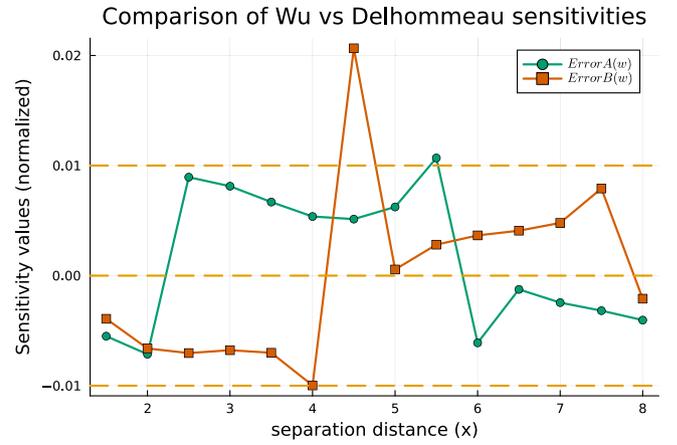

**Fig. B.17.** Error in added mass and damping sensitivities using Wu's (Wu et al., 2017) approximation vs Delhommeau's (Delhommeau, 1987) exact Green's function at $\omega = 1.03$ rad/s.

and

$$K\sigma = \frac{\partial\Phi}{\partial n} \tag{A.8}$$

where $K$ is the adjoint double-layer operator defined as

$$K_{ij} = \frac{\delta_{ij}}{2} + \iint_{S_{B_i}} \nabla_1 G(x_i,\xi)\cdot n_i d\xi \tag{A.9}$$

where $n_i$ denotes the normal vector of panel $S_{B_i}$.

### A.3. Relationship between double-layer operators

The relationship between the gradients $\nabla_1 G$ and $\nabla_2 G$ is derived from the symmetry property of the Green's function:

$$G(x,\xi) = G(\xi,x). \tag{A.10}$$

Differentiating with respect to $\xi$, we have:

$$\nabla_2 G(x,\xi) = \nabla_1 G(\xi,x). \tag{A.11}$$

This symmetry allows for efficient computation of $D$ and $K$ using shared code, with minor adjustments for switching between symmetric and antisymmetric components.

## Appendix B. Comparison of sensitivities using wu and delhommeau's approximation

Here we compare the hydrodynamic sensitivities computed using the Green's function approximation-based solvers of Wu (Wu et al., 2017) and Delhommeau (Delhommeau, 1987), with respect to the separation distance between the spheres described in Section 4.1.

This comparison verifies that the numerical approximation by Wu et al. (Wu et al., 2017) for estimating gradient sensitivities aligns within 2% of the exact expressions derived by Delhommeau (Delhommeau, 1987). Since both expressions are available, either can be used, but this comparison confirms that the faster Wu approximation is also reliable for gradient estimation, making it a viable option for computational efficiency. This highlights the trade-off between using the fast, approximate Green's function and its gradients versus the exact but slower Green's function and its derivatives.

## Appendix C. Comparison of hydrodynamic sensitivities for direct and indirect BIE formulation

Hydrodynamic sensitivities computed via direct and indirect BIE formulation using surrogate Green's function.

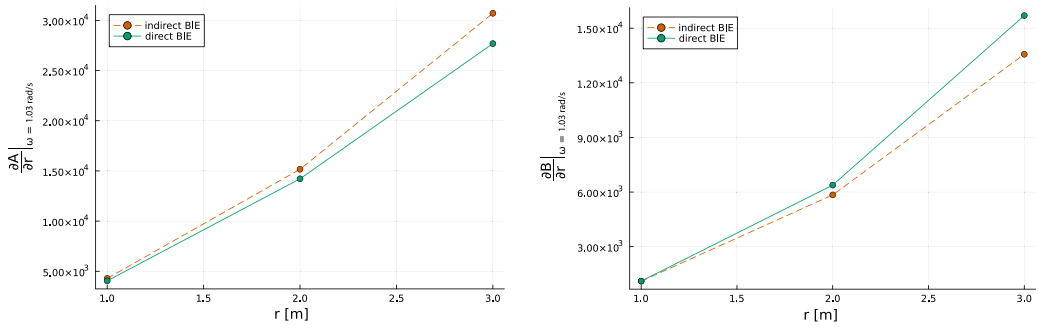

**Fig. C.18.** Comparison of gradients computed by AD for direct and indirect BIE formulations for (a) heave added mass $\frac{\partial A}{\partial r}\big|_{\omega=1.03 \text{ rad/s}}$ (b) and heave damping $\frac{\partial B}{\partial r}\big|_{\omega=1.03 \text{ rad/s}}$.

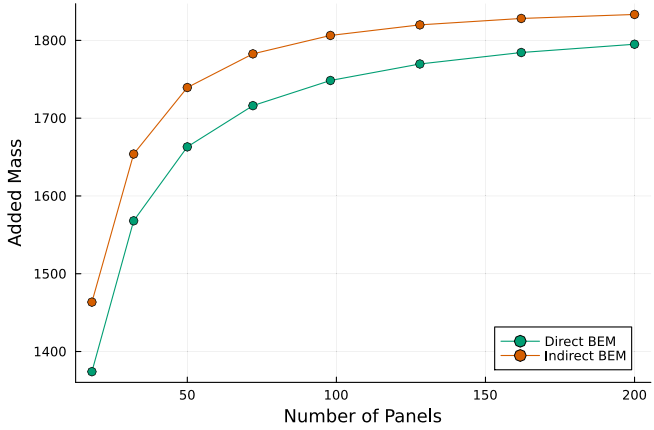

**Fig. E.19.** Mesh convergence study for both direct and indirect BEM formulations.

## Appendix D. Custom rules for geometry and hydrostatics calculations

See Algorithm 1.

---

**Algorithm 1** Custom reverse-mode differentiation rule (rrule) for mesh computations $f_{\text{mesh}}(x)$

---
1: **Input:** $x$ (input vector), $dy$ (perturbation from the output)
2: **Output:** $y$ (function output), $dx$ (gradient w.r.t. $x$)
3: Define $f_{\text{mesh}}(x)$ external operation:
4: **function** RRULE($f_{\text{mesh}}$, $x$)
5:      $y \leftarrow f_{\text{mesh}}(x)$       ▷ Compute the primal value
6:      **function** PULLBACK(dy)
7:          $df \leftarrow \text{NoTangent}()$     ▷ No tangent contribution
8:          $dx \leftarrow \frac{\partial f_{\text{mesh}}}{\partial x}^T \cdot \delta y$   ▷ Compute Vector-Jacobian product
9:          **return** $(df, dx)$
10:      **return** $(y, \text{pullback})$

---

## Appendix E. Mesh convergence

The mesh convergence study in Fig. E.19 shows that more than 98 panels are needed to ensure convergence for the coefficients for both the indirect and direct BIE methods. The figure also shows that the two methods have slightly different convergence behavior and converge to different values. It is often mentioned in the literature that these two formulations differ in their behavior, but the cause is not well explained, although theoretically they should be equivalent (Papillon et al., 2020).

## Appendix F. Adjoint via method of Lagrange multipliers

Consider minimizing $\mathcal{J}(\phi, \theta)$ with respect to $\phi$ and $\theta$, where $\phi(\theta)$ is defined implicitly by the equation

$$D(\theta)\phi(\theta) - S(\theta)b(\theta) = 0.$$

The minimizer will be a critical point of the Lagrangian

$$\mathcal{L}(\phi, \theta, \lambda) = \mathcal{J}(\phi, \theta) - \lambda^T \left( D(\theta)\phi - S(\theta)b(\theta) \right),$$

which has derivatives

$$\frac{\partial \mathcal{L}}{\partial \phi} = \frac{\partial \mathcal{J}}{\partial \phi} - \lambda^T D(\theta)$$

$$\frac{\partial \mathcal{L}}{\partial \theta} = \frac{\partial \mathcal{J}}{\partial \theta} + \lambda^T \left( S \frac{\partial b}{\partial \theta} + \frac{\partial S}{\partial \theta} b - \frac{\partial D}{\partial \theta} \phi \right)$$

$$\frac{\partial \mathcal{L}}{\partial \lambda} = (D\phi - Sb)^T.$$

These equations define the necessary conditions for a critical point of the Lagrangian. Along the constraint set, the derivative of $\mathcal{J}$ with respect to $\theta$ is given by $\frac{\partial \mathcal{L}}{\partial \theta}$, where the Lagrange multipliers ($\lambda$) can be computed by setting $\frac{\partial \mathcal{L}}{\partial \phi}$ to zero.

## Appendix G. Optimization step by step: Adjoint vs finite difference

---

**Algorithm 2** Adjoint Method for Optimization (using differentiable solver)

---
1. **Input:** Initial guess for design variables, $\theta_0$
2. **Output:** Optimal design variables $\theta^*$
3. Solve the forward diffraction and/or radiation problem to get the objective function $f(\theta)$
4. Compute the adjoint variables $\lambda$ based on the forward problem solution
5. Compute the gradient of the objective function with respect to optimization variables using adjoint variable
6. Update design variables using a gradient-based optimization algorithm.
7. Repeat steps 3 to 5 until convergence within tolerance

---

**Algorithm 3** Finite Difference Method for Optimization (naive way using existing BEM solvers)

---
1. **Input:** Initial guess for optimization variables, $\theta_0$
2. **Output:** Optimal variables $\theta^*$
3. Solve the forward diffraction and/or radiation problem to get the objective function $f(\theta)$
4. **for** each design variable $\theta_i$

    (a) Perturb $\theta_i$ by a small amount $\epsilon$: $\theta_i^+ = \theta_i + \epsilon$
    (b) Solve the forward problem again with perturbed variable: $f(\theta^+)$
    (c) Compute the finite difference gradient:

$$\frac{\partial f}{\partial \theta_i} \approx \frac{f(\theta^+) - f(\theta)}{\epsilon}$$

5. Update design variables using the computed gradients
6. Repeat steps 3 to 6 until convergence within tolerance

---

## References

Allaire, Grégoire, 2015. A review of adjoint methods for sensitivity analysis, uncertainty quantification and optimization in numerical codes. Ing. Automob. 836.

Ancellin, Matthieu, 2024. On the accuracy of linear boundary-element-method sea-keeping codes. In: 19e Journées de l'Hydrodynamique, École Centrale de Nantes. Nantes, France.

Ancellin, Matthieu, Dias, Frédéric, 2019. Capytaine: a Python-based linear potential flow solver. J. Open Source Softw. 4 (36), 1341.

Ashuri, T., Zaaijer, M.B., Martins, J.R.R.A., van Bussel, G.J.W., van Kuik, G.A.M., 2014. Multidisciplinary design optimization of offshore wind turbines for minimum levelized cost of energy. Renew. Energy 68, 893–905.

Babarit, Aurélien, Delhommeau, Gérard, 2015. Theoretical and numerical aspects of the open source BEM solver NEMOH. In: Proceedings of the 11th European Wave and Tidal Energy Conference. EWTEC2015, Nantes, France.

Bartholomew-Biggs, Michael, Brown, Steven, Christianson, Bruce, Dixon, Laurence, 2000. Automatic differentiation of algorithms. J. Comput. Appl. Math. 124 (1), 171–190.

Bezanson, Jeff, Edelman, Alan, Karpinski, Stefan, Shah, Viral B, 2017. Julia: a fresh approach to numerical computing. SIAM Review 59 (1), 65–98.

Bigoni, Davide, 2015. Uncertainty Quantification with Applications to Engineering Problems (Ph.D. thesis). Technical University of Denmark.

Blondel, Mathieu, Roulet, Vincent, 2024. The elements of differentiable programming.

Borgarino, Bruno, Babarit, Aurélien, Ferrant, Pierre, 2012. Impact of wave interactions effects on energy absorption in large arrays of wave energy converters. Ocean Eng. 41, 79–88.

Bradley, Andrew M., 2010. PDE-constrained optimization and the adjoint method. original November 16, 2010, revised July 7, 2024.

Burton, A.J., Miller, GF495032, 1971. The application of integral equation methods to the numerical solution of some exterior boundary-value problems. Proc. R. Soc. A 323 (1553), 201–210.

Delhommeau, Gérard, 1987. Problèmes de diffraction-radiation et de résistance des vagues : étude théorique et résolution numérique par la méthode des singularités (Ph.D. thesis). École Nationale Supérieure de Mécanique de Nantes, Nantes, France.

Dou, Suguang, Pegalajar-Jurado, Antonio, Wang, Shaofeng, Bredmose, Henrik, Stolpe, Mathias, 2020. Optimization of floating wind turbine support structures using frequency-domain analysis and analytical gradients. J. Phys.: Conf. Ser. 1618 (4), 042028.

Elliott, Conal, 2018. The simple essence of automatic differentiation. Proc. ACM Program. Lang. 2 (ICFP).

Falnes, Johannes, 2002. Ocean Waves and Oscillating Systems: Linear Interactions Including Wave-Energy Extraction. Cambridge University Press.

Forrester, Alexander I.J., Keane, Andy J., 2009. Recent advances in surrogate-based optimization. Prog. Aerosp. Sci. 45 (1), 50–79.

Gaul, Lothar, Koegl, Martin, Wagner, Marcus, 2003. Boundary element methods for engineers and scientists: An introductory course with advanced topics. 57, p. B31, Springer.

Giles, Michael B., Pierce, Niles A., 2000. An introduction to the adjoint approach to design. Flow, Turbul. Combust. 65, 393–415, Received 13 December 1999; accepted in revised form 2 February 2000.

Gomes, R.P.F., Henriques, J.C.C., Gato, L.M.C., Falcão, A.F.O., 2012. Hydrodynamic optimization of an axisymmetric floating oscillating water column for wave energy conversion. Renew. Energy 44, 328–339.

Gray, Justin S, Hwang, John T, Martins, Joaquim RRA, Moore, Kenneth T, Naylor, Bret A, 2019. OpenMDAO: An open-source framework for multidisciplinary design, analysis, and optimization. Struct. Multidiscip. Optim. 59, 1075–1104.

Griewank, Andreas, 2003. A mathematical view of automatic differentiation. Acta Numer. 12, 321–398.

Griewank, Andreas, Walther, Andrea, 2008. Evaluating Derivatives, second ed. Society for Industrial and Applied Mathematics.

Grund, F., 1985. Pironneau, o., optimal shape design for elliptic systems. ZAMM - J. Appl. Math. Mech. 65 (10), 523–523.

Gunzburger, Max D., 1987. Perspectives in flow control and optimization.

Hascoet, Laurent, Pascual, Valérie, 2013. The tapenade automatic differentiation tool: Principles, model, and specification. ACM Trans. Math. Software 39 (3).

Ho Choi, Joo, Man Kwak, Byung, 1990. A unified approach for adjoint and direct method in shape design sensitivity analysis using boundary integral formulation. Eng. Anal. Bound. Elem. 7 (1), 39–45.

Hulme, A., 1982. The wave forces acting on a floating hemisphere undergoing forced periodic oscillations. J. Fluid Mech. 121 (-1), 443.

Innes, Michael, 2018. Don't unroll adjoint: Differentiating SSA-form programs. CoRR, abs/1810.07951.

Jameson, Antony, 1988. Aerodynamic design via control theory. J. Sci. Comput. 3 (3), 233–260.

John, Fritz, 1950. On the motion of floating bodies II. Simple harmonic motions. Comm. Pure Appl. Math. 3 (1).

Kashiwagi, Masashi, Endo, Kazuaki, Yamaguchi, Hiroshi, 2005. Wave drift forces and moments on two ships arranged side by side in waves. Ocean Eng. 32 (5), 529–555.

Kenway, Gaetan K.W., Mader, Charles A., He, Ping, Martins, Joaquim R.R.A., 2019. Effective adjoint approaches for computational fluid dynamics. Prog. Aerosp. Sci. 110, 100542.

Koh, Chang-Seop, Hahn, Song-Yop, Chung, Tae-Kyung, Jung, Hyun-Kyo, 1992. A sensitivity analysis using boundary element method for shape optimization of electromagnetic devices. IEEE Trans. Magn. 28 (2), 1577–1580.

Konispoliatis, D.N., Mavrakos, S.A., 2016. Hydrodynamic analysis of an array of interacting free-floating oscillating water column (OWCs) devices. Ocean Eng. 111, 179–197.

Lee, C.-H., Newman, J. N., 2003. Computation of wave effects using the panel method, WAMIT Inc., USA (www.wamit.com); Department of Ocean Engineering, MIT, USA.

Lee, C.-H., Sclavounos, P.D., 1989. Removing the irregular frequencies from integral equations in wave-body interactions. J. Fluid Mech. 207, 393–418.

Lettermann, Leon, Jurado, Alejandro, Betz, Timo, Wörgötter, Florentin, Herzog, Sebastian, 2024. Tutorial: a beginner's guide to building a representative model of dynamical systems using the adjoint method. Commun. Phys. 7 (1), 128.

Liang, Hui, Shao, Yanlin, Chen, Jikang, 2021. Higher-order derivatives of the green function in hyper-singular integral equations. Eur. J. Mech. B Fluids 86, 223–230.

Liang, Hui, Wu, Huiyu, Noblesse, Francis, 2018. Validation of a global approximation for wave diffraction-radiation in deep water. Appl. Ocean Res. 74, 80–86.

Liu, Yingyi, 2019. HAMS: A frequency-domain preprocessor for wave-structure interactions—Theory, development, and application. J. Mar. Sci. Eng. 7 (3).

Liu, Yujie, Falzarano, Jeffrey M., 2017. Irregular frequency removal methods: theory and applications in hydrodynamics. Mar. Syst. Ocean. Technol. 12 (2), 49–64.

Mackay, Ed, 2019. Consistent expressions for the free-surface green function in finite water depth. Appl. Ocean Res. 93, 101965.

Mader, Charles A., Martins, Joaquim R.R.A., Alonso, Juan J., van der Weide, Edwin, 2008. Adjoint: An approach for the rapid development of discrete adjoint solvers. AIAA J. 46 (4), 863–873.

Martins, J.R.R.A., Kennedy, G.J., 2021. Enabling large-scale multidisciplinary design optimization through adjoint sensitivity analysis. Struct. Multidiscip. Optim. 64 (5), 2959–2974.

Martins, Joaquim R.R.A., Ning, Andrew, 2022. Engineering Design Optimization. Cambridge University Press.

Mogensen, Patrick Kofod, White, John Myles, Riseth, Asbjørn Nilsen, Holy, Tim, Lubin, Miles, von Salis, Christof, Noack, Andreas, Levitt, Antoine, Ortner, Christoph, Legat, Benoît, Johnson, Blake, Rackauckas, Christopher, Yu, Yichao, Carlsson, Kristoffer, Lin, Dahua, Strouwen, Arno, Grawitter, Josua, Arakaki, Takafumi, Pasquier, Benoît, Covert, Thomas R., Rock, Ron, Creel, Michael, cossio, Regier, Jeffrey, Kuhn, Ben, Stukalov, Alexey, Williams, Alex, Sato, Kenta, 2024. Julianlsolvers/Optim.jl: v1.10.0.

Nadarajah, Siva, Jameson, Antony, 2000. A comparison of the continuous and discrete adjoint approach to automatic aerodynamic optimization.

Newman, J.N., 1984. Approximations for the bessel and struve functions. Math. Comp. 43 (168), 551–556.

Newman, John Nicholas, 1985. Algorithms for the free-surface green function. J. Engrg. Math. 19.

Newman, J.N., 1986. Distributions of sources and normal dipoles over a quadrilateral panel. J. Engrg. Math. 20 (2), 113–126.

Ning, Andrew, McDonnell, Taylor, 2023. Automating steady and unsteady adjoints: Efficiently unifying implicit and algorithmic differentiation. p. 12, 3 figures, arXiv preprint arXiv:2306.15243.

Organization, JuliaDiff, 2024. The propagators: pushforward and pullback.

Papillon, Louis, Costello, Ronan, Ringwood, John V., 2020. Boundary element and integral methods in potential flow theory: a review with a focus on wave energy applications. J. Ocean. Eng. Mar. Energy 6 (3), 303–337.

Patryniak, Katarzyna, Collu, Maurizio, Coraddu, Andrea, 2022. Multidisciplinary design analysis and optimisation frameworks for floating offshore wind turbines: State of the art. Ocean Eng. 251, 111002.

Ragab, Saad A., 2004. Shape optimization of surface ships in potential flow using an adjoint formulation. AIAA J. 42 (2), 296–304.

Ringwood, John V., 2025. Control co-design for wave energy systems. Appl. Ocean Res. 158, 104514.

Rohrer, Peter J., Bachynski-Polić, Erin E., 2024. Analytical gradients of first-order diffraction and radiation forces for design optimization of floating structures. Appl. Ocean Res. 152, 104198.

Ruehl, Kelley, Leon, Jorge, Michelen, Carlos, Topper, Mathew, Tom, Nathan, Baca, Elena, Ogden, David, 2023. Next-Generation Marine Energy Software Needs Assessment. Technical Report, Sandia National Laboratories and National Renewable Energy Laboratory, SAND2023-03906R, NREL/TP-5700-84936.

Rumelhart, David E., Hinton, Geoffrey E., Williams, Ronald J., 1986. Learning representations by back-propagating errors. Nature 323 (6088), 533–536.

Sauter, Stefan A., Schwab, Christoph, 2011. Boundary element methods. Springer Berlin Heidelberg, Berlin, Heidelberg, pp. 183–287.

SciML, 2025. Integrals.jl: Numerical integration tools in julia. (Accessed 08 January 2025).

Silva, Olavo M., Serafim, Luisa P., Mareze, Paulo H., Fonseca, William D'Andrea, Cardoso, Eduardo L., 2023. Discrete approach for shape optimization of 2D time-harmonic acoustic radiation problems solved by BEM using the fully-analytical adjoint method. Eng. Anal. Bound. Elem. 156, 548–571.

Singh, J., Babarit, A., 2013. Hydrodynamic interactions in multiple body array: A simple and fast approach coupling boundary element method and plane wave approximation. In: Proceedings of the International Conference on Offshore Mechanics and Arctic Engineering.

Singh, Jitendra, Babarit, Aurélien, 2014. A fast approach coupling boundary element method and plane wave approximation for wave interaction analysis in sparse arrays of wave energy converters. Ocean Eng. 85 (0029–8018), 12–20.

Sobol', I.M., 2001. Global sensitivity indices for nonlinear mathematical models and their Monte Carlo estimates. Math. Comput. Simulation 55 (1–3), 271–280.

Takahashi, Toru, Sato, Daisuke, Isakari, Hiroshi, Matsumoto, Toshiro, 2022. A shape optimisation with the isogeometric boundary element method and adjoint variable method for the three-dimensional Helmholtz equation. Computer-Aided Des. 142, 103126.

Teixeira-Duarte, Felipe, Clemente, Daniel, Giannini, Gianmaria, Rosa-Santos, Paulo, Taveira-Pinto, Francisco, 2022. Review on layout optimization strategies of offshore parks for wave energy converters. Renew. Sustain. Energy Rev. 163, 112513.

Tortorelli, D.A., Michaleris, P., 1994. Design sensitivity analysis: Overview and review. Inverse Probl. Eng. 1 (1), 71–105.

Towara, Markus, Naumann, Uwe, 2013. A discrete adjoint model for OpenFOAM. Procedia Comput. Sci. 18, 429–438.

White, Frames, Abbott, Michael, Zgubic, Miha, Revels, Jarrett, Robinson, Nick, Arslan, Alex, Widmann, David, Schaub, Simeon David, Ma, Yingbo, Tebbutt, Will, Axen, Seth, Rackauckas, Christopher, Vertechi, Pietro, BSnelling, Fischer, Keno, st, Horikawa, Yuto, Cottier, Ben, Ranocha, Hendrik, Monticone, Pietro, 2024. JuliaDiff/ChainRulesCore.jl: v1.25.0.

Wu, Huiyu, Zhang, Chenliang, Zhu, Yi, Li, Wei, Wan, Decheng, Noblesse, Francis, 2017. A global approximation to the green function for diffraction radiation of water waves. Eur. J. Mech. B Fluids 65, 54–64.

Xie, Chunmei, Choi, Youngmyung, Rongere, François, Clement, Alain H., Delhommeau, Gerard, Babarit, Aurelien, 2018. Comparison of existing methods for the calculation of the infinite water depth free-surface green function for the wave-structure interaction problem. Appl. Ocean Res. 81, 150–163.

Zhang, Jize, Taflanidis, Alexandros A., Scruggs, Jeffrey T., 2020. Surrogate modeling of hydrodynamic forces between multiple floating bodies through a hierarchical interaction decomposition. J. Comput. Phys. 408, 109298.

Zhang, Deqing, Yuan, Zhi-Ming, Du, Junfeng, Li, Huajun, 2022. Hydrodynamic modelling of large arrays of modularized floating structures with independent oscillations. Appl. Ocean Res. 129, 103371.

Zheng, S., et al., 2024. Recent advances in marine hydrodynamics. Phys. Fluids 36 (7), 070402.